\documentclass{jfm2}

\usepackage{amssymb,graphics,color,mathrsfs}
\usepackage{amsmath}
\usepackage{bm}
\usepackage{graphicx}
\usepackage{natbib}

\numberwithin{equation}{section}

\usepackage{color}
\definecolor{red}{rgb}{1.0,0.0,0.0}
\definecolor{blue}{rgb}{0.0,0.0,1.0}

\newcommand{\degrees}{\ensuremath{^\circ}}

\title[Cilia-driven flow in zebrafish]{Symmetry breaking cilia-driven flow in the zebrafish embryo}
\author[A.A. Smith \& T.D. Johnson {\it et al.}]{ANDREW A. SMITH$^{1,2}$\thanks{The first and second authors contributed equally to this work}, THOMAS D. JOHNSON$^{1,2}\dagger$, DAVID J. SMITH$^{1,2,3}$\thanks{Email address for correspondence: d.j.smith.2@bham.ac.uk} \and JOHN R. BLAKE$^{1,2}$}
\affiliation{$^1$School of Mathematics, University of Birmingham, Edgbaston, Birmingham, B15 2TT, UK\\[\affilskip]
$^2$Centre for Human Reproductive Science, Birmingham Women's NHS Foundation Trust, Edgbaston, Birmingham, B15 2TG, UK\\[\affilskip]
$^3$School of Engineering \& Centre for Scientific Computing, University of Warwick, Coventry, CV4 7AL, UK}
\date{}

\begin{document}
\maketitle

\section*{Abstract}
\label{abstract}

Fluid mechanics plays a vital role in early vertebrate embryo development, an example being the establishment of left-right asymmetry. Following the dorsal-ventral and anterior-posterior axes, the left-right axis is the last to be established; in several species it has been shown that an important process involved with this is the production of a left-right asymmetric flow driven by `whirling' cilia. It has previously been established in experimental and mathematical models of the mouse ventral node that the combination of a consistent rotational direction and posterior tilt creates left-right asymmetric flow. The zebrafish organising structure, Kupffer's vesicle, has a more complex internal arrangement of cilia than the mouse ventral node; experimental studies show the flow exhibits an anticlockwise rotational motion when viewing the embryo from the dorsal roof, looking in the ventral direction. Reports of the arrangement and configuration of cilia suggest two possible mechanisms for the generation of this flow from existing axis information: (1) posterior tilt combined with increased cilia density on the dorsal roof, and (2) dorsal tilt of `equatorial' cilia. We develop a mathematical model of symmetry breaking cilia-driven flow in Kupffer's vesicle using the regularized Stokeslet boundary element method. Computations of the flow produced by tilted whirling cilia in an enclosed domain suggest that a possible mechanism capable of producing the flow field with qualitative and quantitative features closest to those observed experimentally is a combination of posteriorly tilted roof and floor cilia, and dorsally tilted equatorial cilia.

\section{Introduction}
\label{sec:intro}

Vertebrates, from the outside, appear bilaterally symmetric. However, in many species, internal body plans are arranged asymmetrically in an organised way. For example the heart can usually be found on the left in mice, zebrafish \citep{Ibanes09}, and humans (figures~\ref{fig:role-cilia}a). As with convention the `left' of the depicted individual is on the right of figure~\ref{fig:role-cilia}a,b and in the majority of figures in this paper.

\begin{figure}%
	\centering
	\caption{Copyright figure available in published copy. Body plans and
axes in humans and zebrafish. (a) Normal organ placement and (b) reversed organ
placement as found in {\it situs inversus}. (c) Schematic outlining how the
biological axes will be represented in Cartesian axes for use in the
mathematical model. (d) A wildtype (WT) zebrafish with body axes overlayed,
reprinted and adapted from \citet{Rawls01}. (e)
Tail view of zebrafish embryo, arrow indicates Kupffer's vesicle, anterior is at
the top and posterior is at the bottom of the figure, reprinted from
\citet{Kreiling07}. Axis notation: {\it d}, dorsal;
{\it v}, ventral; {\it a}, anterior; {\it p}, posterior; {\it l}, left; and {\it
r}, right.}
	\label{fig:role-cilia}
\end{figure}

Three axes are established during early embryonic development; in order they are the dorsal-ventral, anterior-posterior and left-right axes (figure \ref{fig:role-cilia}c) \citep{Hirokawa09}. It is only relatively recently that the mechanisms involved in the initial establishment of left-right asymmetry have begun to be understood. \citet{Kartagener33}, amongst others, identified a triad of conditions that included respiratory problems, male infertility and {\it situs inversus}, the lateral transposition of internal organs (figure \ref{fig:role-cilia}b). For a detailed historical review see also \citet{Berdon04b} and \citet{Berdon04a}. In $1974$, Afzelius performed electron microscopy on the sperm of four infertile men. The flagella, which were immotile, did not possess the dynein motor protein. Three of the four men did not have normally-functioning lung cilia, moreover three of the four men were reported to have {\it situs inversus}. From this evidence, Afzelius hypothesised that, ``Visceral asymmetry is determined through the movements of cilia of some embryonic epithelial tissues'' \citep{Afzelius76}.

Twenty years later, \citet{Sulik94} discovered a node structure on a mouse embryo at $7$--$9$ days post-fertilisation that expressed primary cilia.
Cilia are microscopic hair-like organelles with an internal arrangement of nine pairs of microtubules arranged in a cylindrical structure. Motile `$9+2$' cilia, similar to the flagella of sperm, possess a central microtubule pair, and perform essential biological functions such as mucus clearance in the lung and transport of the ovum in the fallopian tube. Primary cilia by contrast are generally shorter, do not possess a central microtubule pair, and prior to the work of \citeauthor{Sulik94} were generally thought to be immotile. However, \citeauthor{Sulik94} reported that the primary cilia in the mouse node were motile and furthermore postulated that their motility may be ``associated with the establishment of sidedness'', thereby providing the link between {\it situs inversus} and ciliary dysfunction in the node. \citet{Nonaka98} then confirmed that primary cilia in the mouse node were indeed motile and in normal mice performed a clockwise rotation, when viewed from tip to base, which propels fluid. \citeauthor{Nonaka98} showed that knockout mice with absent cilia do not break symmetry normally and then do not survive gestation.
\citet{Nonaka02} subsequently showed that a leftward fluid flow was both necessary and sufficient for normal organ {\it situs} by placing embryos in artificial flow conditions showing that a rightward flow resulted in {\it situs inversus}. However at the time of \citeauthor{Nonaka02}'s work it was still unclear as to how a whirling motion could be sufficient to create a directional fluid flow.

%There have been many advances in both biological and mathematical studies relating to cilia-driven fluid flows.
A resolution to this question was first proposed by \citet{Cartwright04}. Representing cilia by  point torque rotlet singularities in an infinite domain, \citeauthor{Cartwright04} showed how a clockwise rotation can generate a leftward fluid flow if the axis of rotation is tilted towards the already-established posterior direction.
This theoretical prediction was verified soon after in biological observations \citep{Okada05} and a mechanical experimental model \citep{Nonaka05}. Moreover, \citeauthor{Cartwright04}'s theoretical prediction that the tilt angle would be approximately $24^\circ$ was close to the mean value of $26.6^\circ$ reported by \citeauthor{Nonaka05}  \citet{Brokaw05a} used a computational model to show how internal chirality of dynein regulation, and internal twist, are possible mechanisms that may create the whirling motion of cilia. Furthermore \citeauthor{Brokaw05a} gave the first discussion of the effect of surface drag in the context of tilted rotational cilia motion, and its effect in generating a leftward flow.

%They then predicted that for a leftward flow that a tilt towards the posterior combined with clockwise rotations was needed.

\citet{Smith07} used a computational approach to slender body theory for Stokes flow to address the roles of cilium-surface interactions and unsteadiness in the flow field, via image systems in a plane boundary \citep{Blake71a}. Evidence of limited chaotic advection was found and later verified experimentally \citep{Supatto08,Supatto11}. \citet{Smith08} also investigated the optimal tilt angle for maximum fluid propulsion by tilted rotational motion, and furthermore modelled particle drift under the action of arrays of tilted whirling cilia; results were found to lie within the bounds reported from experimental observations \citep{Okada05,Nonaka05}, and also the mechanical analogue model of \citet{Nonaka05}. Addressing the issue that the developing node has an enclosing upper membrane,
\citet{Cartwright07} computed flow profiles using a steady flow finite element model. It was predicted that there would exist two layers of rightward `return' flow, in the `upper' region away from the cilia, caused by conservation of mass in the volume, and also very close to the ciliated surface, the latter caused by the return stroke of the cilia.
To investigate the influence of the enclosed fluid domain in the context of a time-dependent model,
\citet{Smith11} used the regularized Stokeslet boundary element method combined with slender body theory. The time-dependent model predicted that the leftward particle drift would occur throughout most of the ciliated layer, including very close to the ciliated surface. It was also found that the main directional flow only extended as far as the cilia array.

An important question is how a directional flow is translated into asymmetric development. Two main models have been proposed: one that the flow transports morphogen proteins to the left of the node, the other is that there are also immotile cilia in organising structures that are there to `sense' the flow produced by the motile cilia. These theories are reviewed and discussed by \citet{Hirokawa09}; a detailed understanding of the flow, transport of particles, and hydrodynamic stresses produced by whirling cilia will be essential to understand this mechanism.

\section{Geometry of organising structures}
\label{sec:geometry}

Many theoretical \citep{Cartwright04,Cartwright07,Smith07,Smith08,Smith11} and early experimental \citep{Nonaka98,Nonaka02,Nonaka05,Okada05,Tanaka05} studies have been conducted based on the geometry of the mouse node. This is because the mouse was one of the first species found to have cilia in its organising structure (see also reviews by \citet{Cartwright08,Cartwright09,Hirokawa09} for full details).

The mouse node is a triangular depression measuring $50$--$100\mu$m in width and $10$--$20\mu$m in depth that forms on the ventral surface of the embryo. The mouse node is covered with a membrane, which is frequently removed for imaging purposes, and is filled with fluid. Primary cilia measuring $3$--$5\mu\textrm{m}$ are expressed on the surface forming the base of the mouse node and they exhibit tilted clockwise rotations as described in the previous section.

In recent years experimentalists have turned attention to imaging the organising structure of the zebrafish (figure \ref{fig:role-cilia}d,e), termed Kupffer's vesicle (KV) \citep{Kawakami05,Kreiling07,Okabe08,Supatto08}. Zebrafish KV is transparent \citep{Supatto11}, meaning that internal structures can be imaged without removing any surfaces. Another distinct advantage of using KV to investigate development is that it is visible at approximately $12$ hours post-fertilisation \citep{Kimmel95}.

\begin{figure}%
	\centering
		\includegraphics[width=4.85in]{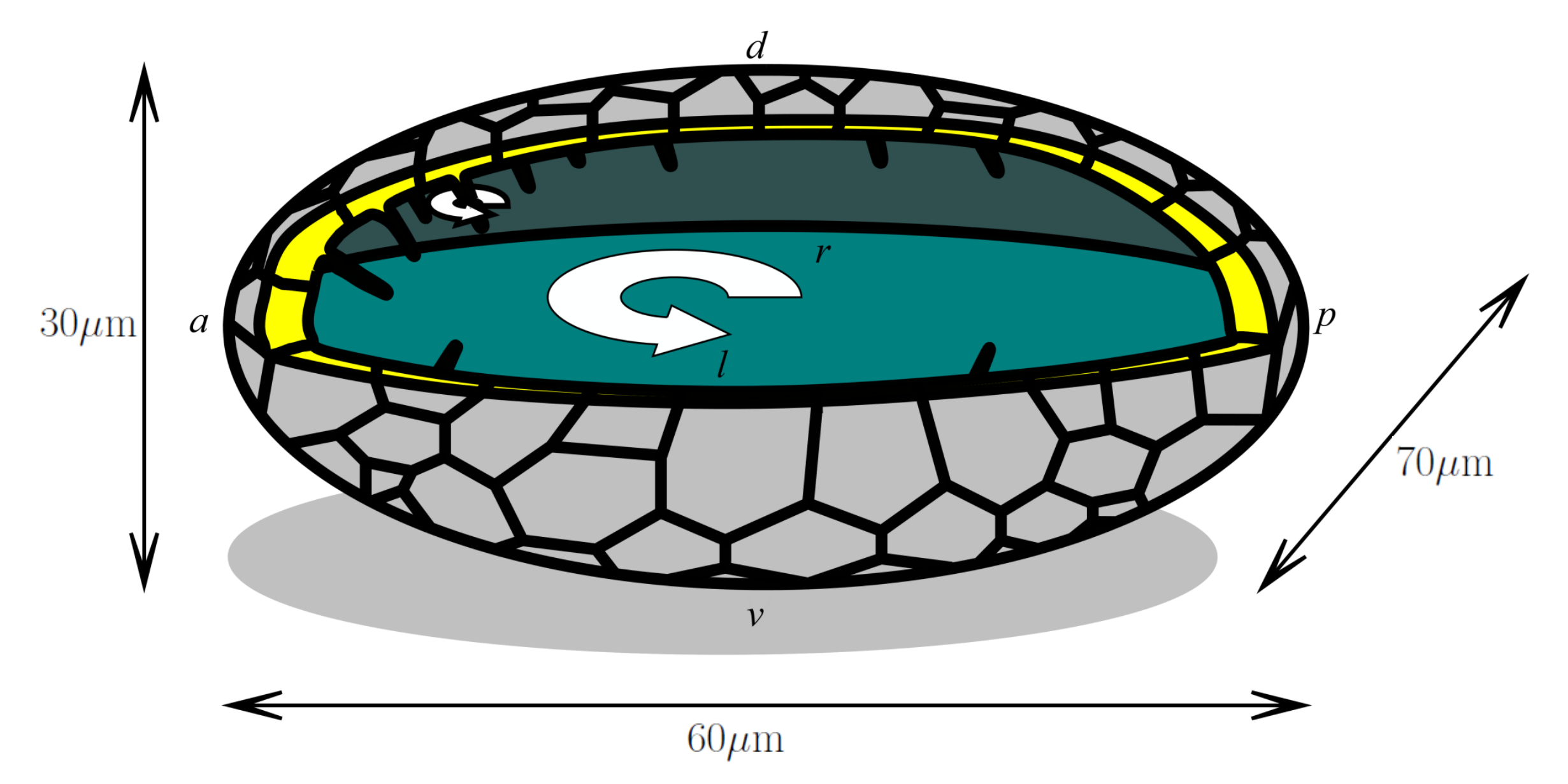}
	\caption{Schematic of the zebrafish Kupffer's vesicle showing cilia positions and their motion. The greater proportion of cilia in the dorsal-anterior region is shown along with the posterior tilt of each cilium. Axis notation: {\it d}, dorsal; {\it v}, ventral; {\it a}, anterior; {\it p}, posterior; {\it l}, left; and {\it r}, right. Figure redrawn from \citet{Kreiling07}.}
	\label{fig:KV}
\end{figure}

Zebrafish KV is a closed spheroidal structure, measuring approximately $70\; \mu\textrm{m}\times60 \; \mu\textrm{m}\times30\; \mu\textrm{m}$, with cilia measuring $2$--$4\; \mu\textrm{m}$ \citep{Kramer05} expressed on the internal surfaces (figure \ref{fig:KV}) \citep{Kreiling07,Supatto08}. It is agreed that the cilia in KV are tilted and rotate, however there is not a consensus as to the tilt direction. \citet{Kreiling07} report that the roof and floor cilia are posteriorly tilted; \citet{Supatto08} report dorsally tilted side wall cilia. Both \citeauthor{Kreiling07} and \citeauthor{Supatto08} describe the same flow field, a circulation about the dorsal-ventral axis (figure \ref{fig:observed_flow}). \citet{Supatto11} suggest that dorsal tilt is necessary to create the observed flow field. Moreover, \citet{Supatto11} report the directional flow magnitude in the region up to approximately $20\; \mu\mathrm{m}$ above the cell surface as being in a range of approximately $6$--$13\; \mu\mathrm{m}/\mathrm{s}$.

\begin{figure}
	\centering
	\caption{Copyright figure available in published copy. Experimental
observations of the flow inside the zebrafish Kupffer's vesicle as reported by
\citet{Supatto08}. The axes
$x$, $y$, $z$ correspond to $x_1$, $x_2$, $x_3$ in the present study. (a)
Measured fluid velocity field around a single cilium, with velocity magnitude
represented by arrow length and colour, extracted from three dimensional
particle tracking, located in the `midplane' region in the black box shown in
(b). The transition between `directional' flow above the cilium and `vortical'
flow around the cilium are indicated; this transition can be understood by
considering the image system of \citet{Blake74a} (see \citet{Smith07} for
further discussion). (c) View from posterior, showing a $30\degrees$ dorsal tilt
of the beating axis. (d) Particle tracks showing the steady-state flow around
the interior of Kupffer's vesicle. The three different track colours correspond
to the superposition of tracks obtained at three $15$ s time windows. The
particles exhibit a circular motion around the dorsal-ventral direction. A three
dimensional schematic representation is shown in (e). The black arrows indicate
the anticlockwise rotation of the flow (when viewed from the dorsal side). Axis
notation: {\it d}, dorsal; {\it v}, ventral; {\it a}, anterior; {\it p},
posterior; {\it l}, left; and {\it r}, right.}
	\label{fig:observed_flow}
\end{figure}

This theoretical study is designed to investigate the flow in KV generated by these models: (1) all posteriorly tilted cilia, (2) all dorsally tilted cilia and (3) a hybrid of posteriorly and dorsally tilted cilia depending on position on the internal surface of KV. Furthermore, we shall investigate the effect of the more densely ciliated anterior-dorsal roof, as also observed in experiment.

In the next section we shall describe the mathematical theory behind our model starting with a description of the physics of a cilium beat cycle. We then formulate the regularized Stokeslet boundary integral equation because it is highly suited to modelling complex geometric problems. We will also describe our method of meshing the interior domain and cilia of KV and then solve the regularized Stokeslet boundary integral equation numerically.

\section{Fluid mechanics modelling}
\label{sec:modelling}

The basic fluid mechanical mechanisms underlying how tilted cilia performing a whirling motion create a directional flow, the nodal flow, have been discussed extensively elsewhere \citep{Smith07,Smith08,Smith11}, based on the theory of \citet{Blake71a} and \citet{Blake74a}. We briefly recapitulate these ideas before describing the computational boundary integral model of KV.

\subsection{Stokes flow driven by cilia}
\label{subsec:stokes-flow}

Due to the small length scales and velocities ($L\approx3\mu\mathrm{m}$, $\omega L\approx2\pi\times25\times3\mu\mathrm{m}/\mathrm{s}$), the Reynolds number of flow near a whirling cilium is approximately $10^{-3}$. An early hypothesis regarding the generation of the nodal flow was based on the speed difference between the leftward and rightward strokes. It was known that a cilium moves more rapidly during the leftward, upright portion of the stroke than during the rightward part where the cilium moves close to the surface. The greater speed of the leftward stroke was hypothesised to produce a greater flow. This observation likely originated from intuitions regarding inertia-dominated high Reynolds number flow, however these intuitions are not accurate in the very low Reynolds number viscous-dominated regime of cilia-driven flow. Very low Reynolds number flows essentially respond `instantaneously' to boundary conditions driving the flow; if the driving velocity is changed, the flow changes proportionately. If there are no other physical effects present, the lower velocity associated with the slower rightward movement is balanced by the fact that the rightward movement occurs for a longer period of time.

The physical effect producing asymmetric flow is `wall interaction'. Walls have very significant effects in Stokes flow, notably converting the $\mathcal{O}(1/r)$ decay of a concentrated force to an $\mathcal{O}(1/r^2)$ decay \citep{Blake71a}. The cilium moves very close to the cell surface during the rightward recovery stroke, but is fairly upright during the leftward effective stroke. As a result the amount of flow in the surrounding fluid produced by the rightward stroke is much smaller than that produced by the leftward stroke. The effect of the cell surface is to reduce the influence of the moving cilium on the bulk of the fluid; this effect is much stronger during the rightward than leftward strokes. The image systems associated with forces acting near walls in Stokes flow give further insight into the nature of the flow field \citep{Blake74a,Smith07}.

The fundamental solution of Stokes flow is the Stokeslet, defined as the solution of the Stokes flow equations driven by a point force. For a force per unit volume $\bm{F}$ located at $\bm{y}$, the Stokes flow equations are,
\begin{equation}
\nabla p = \mu\nabla^2\bm{u} + \bm{F}\delta(\bm{x} - \bm{y})\text{,}\quad \nabla\cdot\bm{u} = 0 \mbox{,}
\label{eq:stokes}
\end{equation}
where $p$ is pressure, $\bm{u}$ is velocity, and $\mu$ is dynamic viscosity. The solution is given by $u_i=F_jS_{ij}$ (with summation convention), where the second rank tensor $S_{ij}$ is given by,
\begin{equation}
S_{ij}(\bm{x},\bm{y}) = \frac{1}{8\pi\mu}\left(\frac{\delta_{ij}}{r} + \frac{r_ir_j}{r^3}\right) \mbox{,}
\label{eq:stokeslet}
\end{equation}
where $r_i = x_i - y_i$ and $r^2 = r_1^2 + r_2^2 + r_3^2$, see for example \citet{Pozrikidis92}.

\citet{Smith07,Smith08,Smith11} used slender body theory to model the cilia-driven flow in the mouse node, which is based on the use of a line distribution of Stokeslets and appropriately-weighted source dipoles. The ciliated surface was modelled using the plane boundary image systems of \citet{Blake71a}. The upper membrane of the mouse node was taken into account \citep{Smith11} using regularized Stokeslets combined with the image system of \citet{Ainley08}.

In this study we wish to account for the curvature of the inner surface of KV, and the relatively low slenderness ratio of 1:10 of the cilia. We shall use the regularized Stokeslet boundary element method, and shall use a surface mesh of the internal cavity, and the cilia themselves.
\citet{Cortez01} introduced the `regularized Stokeslet'. This is defined as the exact solution to the Stokes flow equations with smoothed point forces,
\begin{equation}
\bm{\nabla}p=\mu\nabla^2\bm{u}+\bm{F}\psi_\epsilon(\bm{x}-\bm{y}) \mbox{,} \quad \bm{\nabla}\cdot\bm{u}   =  0        \mbox{.}
\end{equation}

\noindent The symbol $\psi_\epsilon(\bm{x}-\bm{y})$ denotes a cutoff-function or `blob' with regularisation parameter $\epsilon$, satisfying $\int_{\mathbb{R}^3} \psi_\epsilon(\bm{x})\mathrm{d}V_{\bm{x}}=1$. \citet{Cortez05} showed that for the blob $\psi_\epsilon(\bm{x}-\bm{y}):=15\epsilon^4/(8\pi\mu r_\epsilon^7)$, the regularized Stokeslet velocity tensor is given by
\begin{equation}
S_{ij}^\epsilon(\bm{x},\bm{y})=\frac{\delta_{ij}(r^2+2\epsilon^2)+r_i r_j}{r_\epsilon^3} \quad \mathrm{where}  \quad r_\epsilon^2=r^2+\epsilon^2.
\label{regSto}
\end{equation}
\citet{Cortez05} also showed that the resulting regularized Stokeslet boundary integral equation for flow bounded by a surface $D$ is,
\begin{equation}
u_j(\bm{y})=\int_D \left[S_{ij}^\epsilon (\bm{x},\bm{y})f_i(\bm{x})-u_{i}(\bm{x})T_{ijk}^\epsilon (\bm{x},\bm{y}) n_k(\bm{x})\right]\mathrm{d}S_{\bm{x}} + \mathcal{O}(\epsilon^2) \mbox{,}
\end{equation}
for a point $\bm{y}$ in the fluid, and where $n_k(\bm{x})$ is a unit surface normal pointing into the fluid. The `single-layer' density $f_i(\bm{x})$ has dimensions of stress, so that $f_i(\bm{x})\mathrm{d}S_{\bm{x}}$ has dimensions of force, in contrast with the localised force per unit volume $\bm{F}$ of equation~\eqref{eq:stokes}. The symbol $T_{ijk}^\epsilon(\bm{x},\bm{y})$ is the stress tensor corresponding to the regularized Stokeslet. Since our flow domain will consist almost exclusively of rigid surfaces, we neglect the `double-layer potential' arising from $T_{ijk}^\epsilon(\bm{x},\bm{y})$. The single-layer potential arising from $S_{ij}^\epsilon(\bm{x},\bm{y})$ is continuous as $\bm{y}$ approaches the surface, hence we have the approximation,
\begin{equation}
u_j(\bm{y})=\int_D S_{ij}^\epsilon (\bm{x},\bm{y})f_i(\bm{x}) \mathrm{d}S_{\bm{x}}+ \mathcal{O}(\epsilon^2)\mbox{,} \quad \mathrm{for} \quad \bm{y}\in D \mbox{.} \label{eq:rsBIE}
\end{equation}

\subsection{Mesh generation and geometric modelling}
\label{subsec:mesh}

As depicted in figure~\ref{fig:distrib}a, the geometry of KV is approximated as the scalene ellipsoidal surface given by
\begin{equation}
\frac{x_1^2}{35^2} + \frac{x_2^2}{30^2} + \frac{x_3^2}{15^2} = 1,
\label{eq:ellipsoid}
\end{equation}
where one length unit corresponds to $1\; \mu$m. We model a vesicle with 70 cilia, one per cell, which is within the observed range of $73.6\pm5.9$ cilia reported by \cite{Kreiling07} for 13 hours post-fertilisation.

\begin{figure}%
	\centering
	\caption{Copyright figure available in published copy. (a) The geometry of Kupffer's vesicle. The red anterior region
has the highest density of cilia, whereas the light blue posterior region has
the lowest, with percentages taken from \citet{Kreiling07}. The cell-structure
indicates the boundaries of the Voronoi cells around each cilium. (b)
Experimentally observed distribution of cilia in Kupffer's vesicle,
reconstructed from confocal microscopy, \citep{Kreiling07}. The image is a two dimensional view of a three dimensional
rendering of the reconstructed cilia; a, m, p denote the anterior, middle and
posterior regions.}
	\label{fig:distrib}
\end{figure}

\begin{figure}%
	\centering
		\includegraphics[width = 0.6\textwidth]{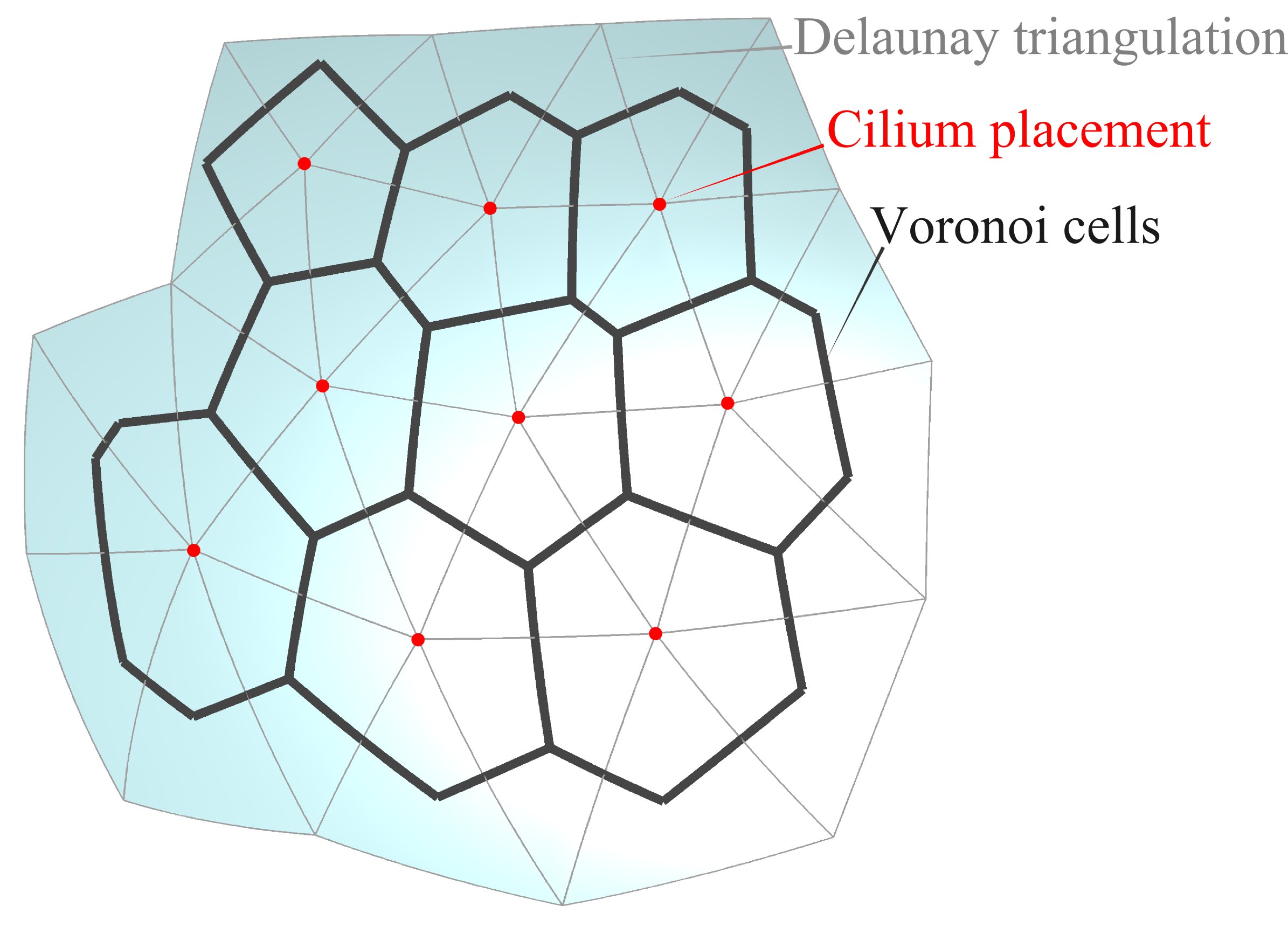}
	\caption{A section of the dorsal roof showing the Delaunay triangulation of the surface, the positions of the cilia, and the cell boundaries created by linking the circumcentres of triangles which share vertices.}
	\label{fig:voronoi_dem}
\end{figure}

\begin{figure}
	\centering
		\includegraphics[width = 0.85\textwidth]{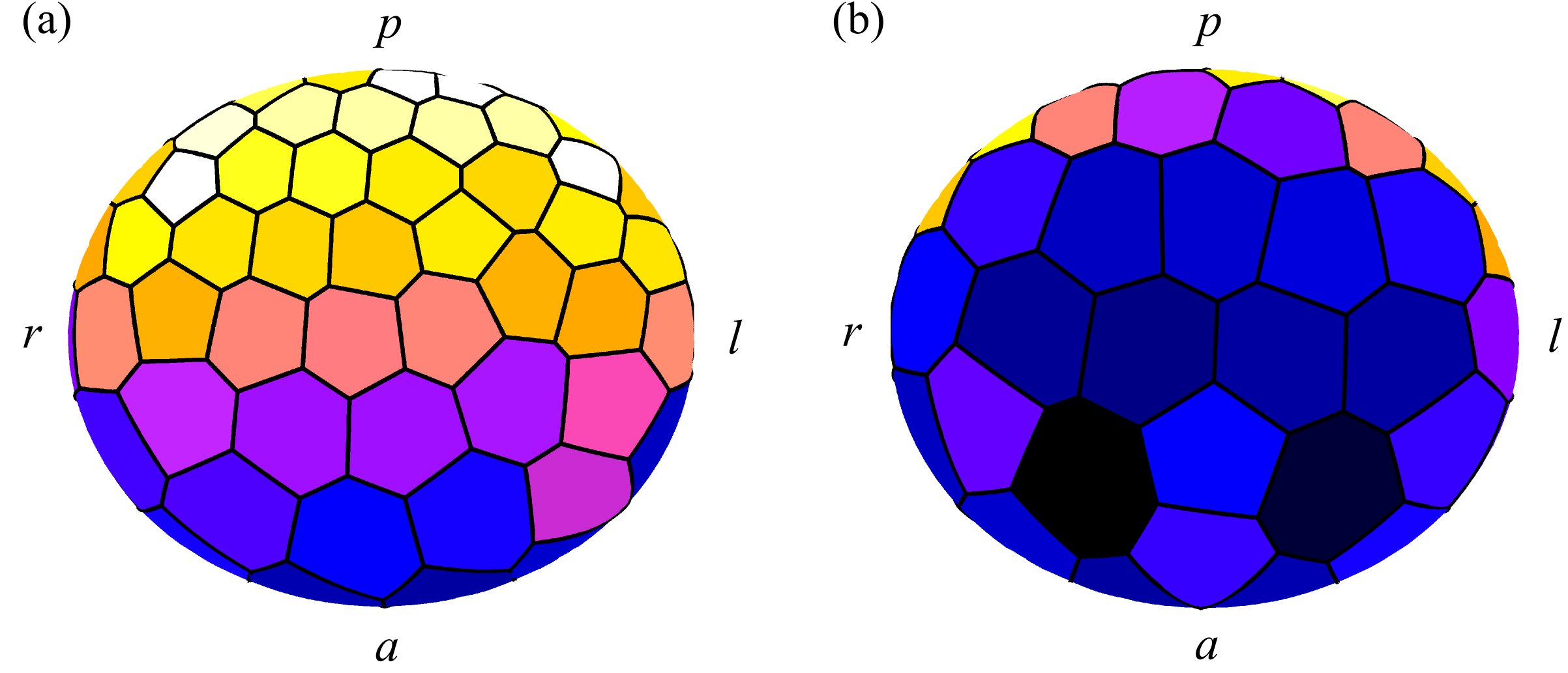}
	\caption{Cell, and hence cilia density, on (a) dorsal and (b) ventral surfaces. Colouring indicates cell surface area, based on figure \ref{fig:distrib}. This cell placement was used to generate the results shown in figure~\ref{fig:results-expt}.}
	\label{fig:dorsal_ventral}
\end{figure}

We wish to test the effect of the experimentally observed distribution of cilia, coupled with cilium tilt, on the production of flow inside KV. To achieve this, we first generate a cell-grid that defines the positions of cilia and the boundaries of each cell by creating a Delaunay surface triangulation using \textit{DistMesh} \citep{Persson04} and custom MATLAB\textregistered \ routines. A triangulation of a set of points is said to have the Delaunay property if the circumcircle of each triangle contains no other points \citep{Okabe92}. A cilium is located at the vertex of each triangle (figure~\ref{fig:voronoi_dem}) so that a triangulation of approximately uniform size yields an even distribution of cilia, whereas a varying triangle size creates a distribution closer to that reported experimentally by \cite{Kreiling07} and shown in figure~\ref{fig:distrib}b. The distribution used is shown in figure~\ref{fig:dorsal_ventral}.
%In order to tesselate the ellipsoidal surface,

The boundaries of each ciliated cell are given by linking the circumcentres of all triangles that share a given vertex, and projecting onto the curved surface defined by equation \eqref{eq:ellipsoid} using the MATLAB\textregistered \ routine \textit{fsolve}. This is an approximation of the intersection of the Voronoi diagram (see also \citeauthor{Okabe92}, \citeyear{Okabe92}) of the set of cilium positions with the ellipsoidal surface, as shown in figure~\ref{fig:distrib}a. The cilium itself comprises a cylindrical section with diameter $0.3\mu$m and length $3\mu$m, and a hemispherical cap.

%\noindent The Voronoi diagram is particularly appropriate for modelling cell-cell contact, as in \cite{schaller2005multicellular}, as it gives the set of points equidistant from any two neighbouring cilia.

\begin{figure}
	\centering
		\includegraphics[width = 0.9\textwidth,viewport=0cm 6.5cm 25.5cm 21cm]{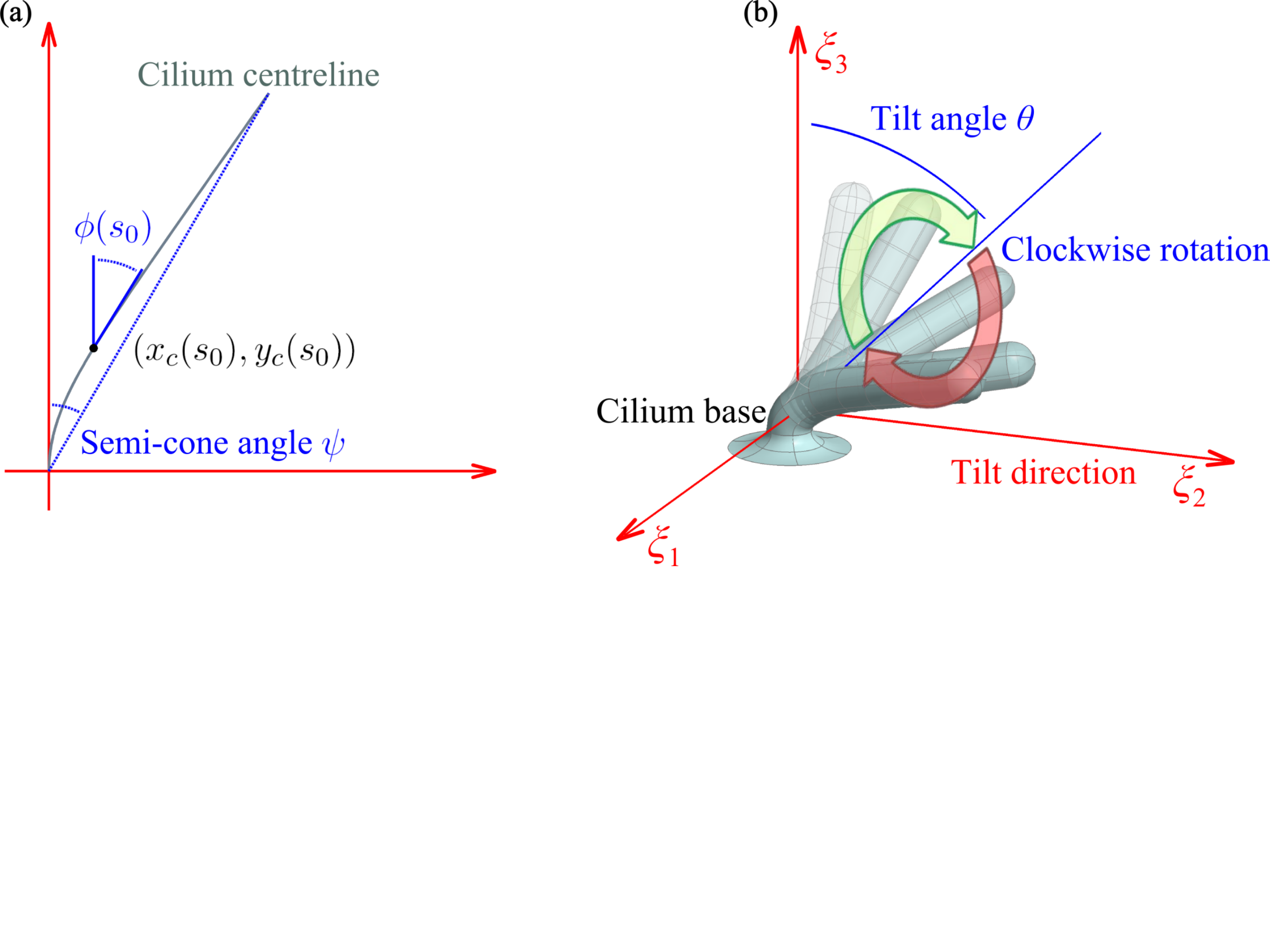}
	\caption{Geometry of the cilium beat. (a) The shape of the cilium centreline about its rotational axis, (b) the final mesh of each cilium showing tilt direction $\xi_2$, tilt angle $\theta$ between the cone axis and wall normal, and the direction of cilia rotation, clockwise when viewed from tip to base. Local coordinates in the cilium frame are denoted $\xi_i$, for $i = 1,2,3$. The upper green arrow shows the effective stroke, the lower red arrow the recovery stroke.}
	\label{fig:cilia}
\end{figure}

The cilia move through an approximately conical envelope with a semi-cone angle $\psi$ of $30\degrees$. The centreline in the cilium frame at time $t$ is given by,
\begin{equation}
\bm{X}(s,t) =
\left(\begin{array}{ccc}
\cos(2\pi t) & \sin(2\pi t) & 0 \\
-\sin(2\pi t) & \cos(2\pi t) & 0 \\
0 & 0 & 1
\end{array}\right)
\left(\begin{array}{c}
0 \\
\int_0^s{\sin(\phi(s^{\prime}))\textrm{d}s^{\prime}} \\
\int_0^s{\cos(\phi(s^{\prime}))\textrm{d}s^{\prime}}
\end{array}\right),
\end{equation}
where $s$ is the arclength along the centreline and $\phi = a \tanh (bs)$. The angle that the tip of the cilium makes with the vertical is given by  $a = 35\degrees$, and the parameter $b = 5$ places the characteristic bend in the base of the cilium shown in figure~\ref{fig:cilia}a. The cilium is attached to the cell smoothly via a fixed base that provides the tilt (figure~\ref{fig:cilia}b).

\begin{figure}%
	\centering
		\includegraphics[width = 0.86\textwidth]{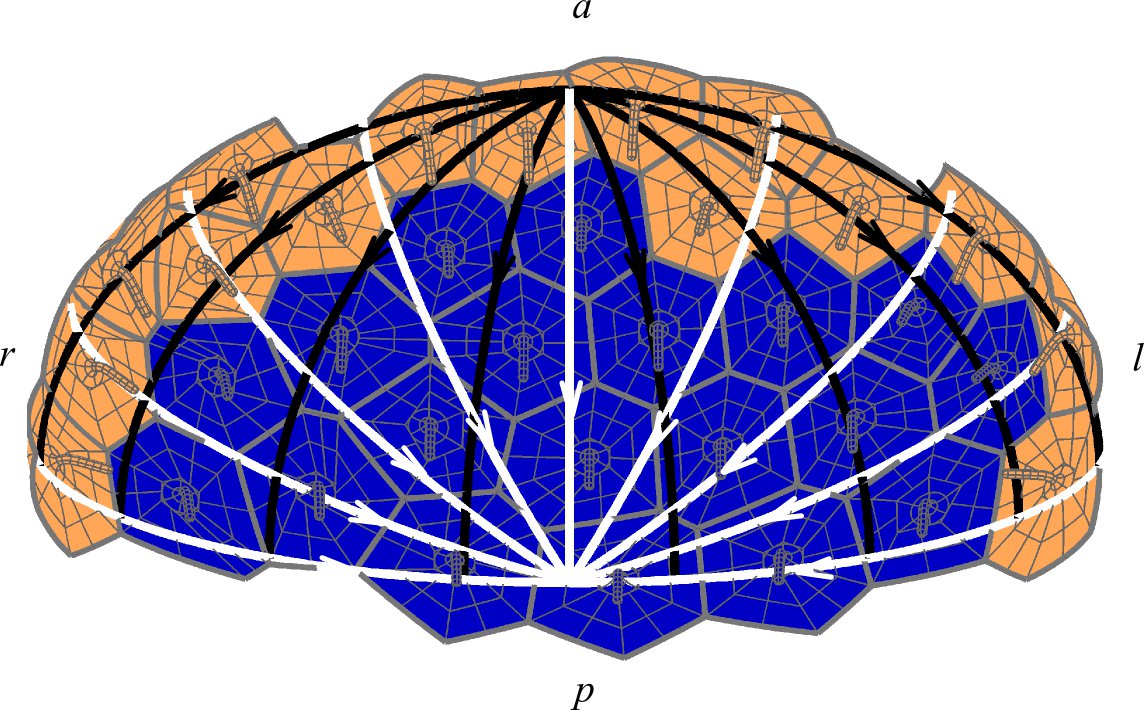}
	\caption{The anterior half of the dorsal roof for one of the meshes used in the computational simulations. The cilium density pattern is chosen to approximate the experimental observations of \citet{Kreiling07}, as shown in figures~\ref{fig:distrib} and \ref{fig:dorsal_ventral}. The point at the top of the figure from which red arrows emanate is the anterior pole; arrows therefore point in the posterior direction. The point at the bottom of the figure on which blue arrows converge is the dorsal pole; arrows therefore point in the dorsal direction. This mesh was generated with equatorial cilia, shaded blue, tilted in the dorsal direction, and dorsal roof cilia, shaded red, tilted in the posterior direction. Later this configuration will be referred to as `mixed tilt'. Results for this mesh are shown later in figures~\ref{fig:results-expt} (g), (h) and (i).}\label{fig:tilt_dirn}
\end{figure}

We consider three separate tilt distributions. Firstly, a purely posterior tilt $\theta=45\degrees$ \citep{Kramer05}, secondly, a purely dorsal tilt of $\theta=30\degrees$ \citep{Supatto08}, and finally, a hybrid model. `Posterior tilt' in our model refers specifically to tilt towards the posterior pole of the vesicle, likewise `dorsal tilt' refers to tilt towards the dorsal pole.
The hybrid model consists of a continuous transition from dorsal tilt close to the `equator' ($x_3=0$) to posterior tilt on the dorsal roof and ventral floor (figure~\ref{fig:tilt_dirn}).

%It is interesting to note that the majority of cilia lie in regions where dorsal tilt and posterior tilt are in roughly the same direction, namely the dorsal-anterior and ventral-posterior corners, so that observed posterior tilt in (Kramer-Zucker) may in fact be dorsal tilt. \newpage

% Cells at the equator with dorsal tilt are coloured blue, whereas the cells at the dorsal pole with posterior tilt are coloured red.

\begin{figure}%
	\centering
		\includegraphics[width = 0.86\textwidth]{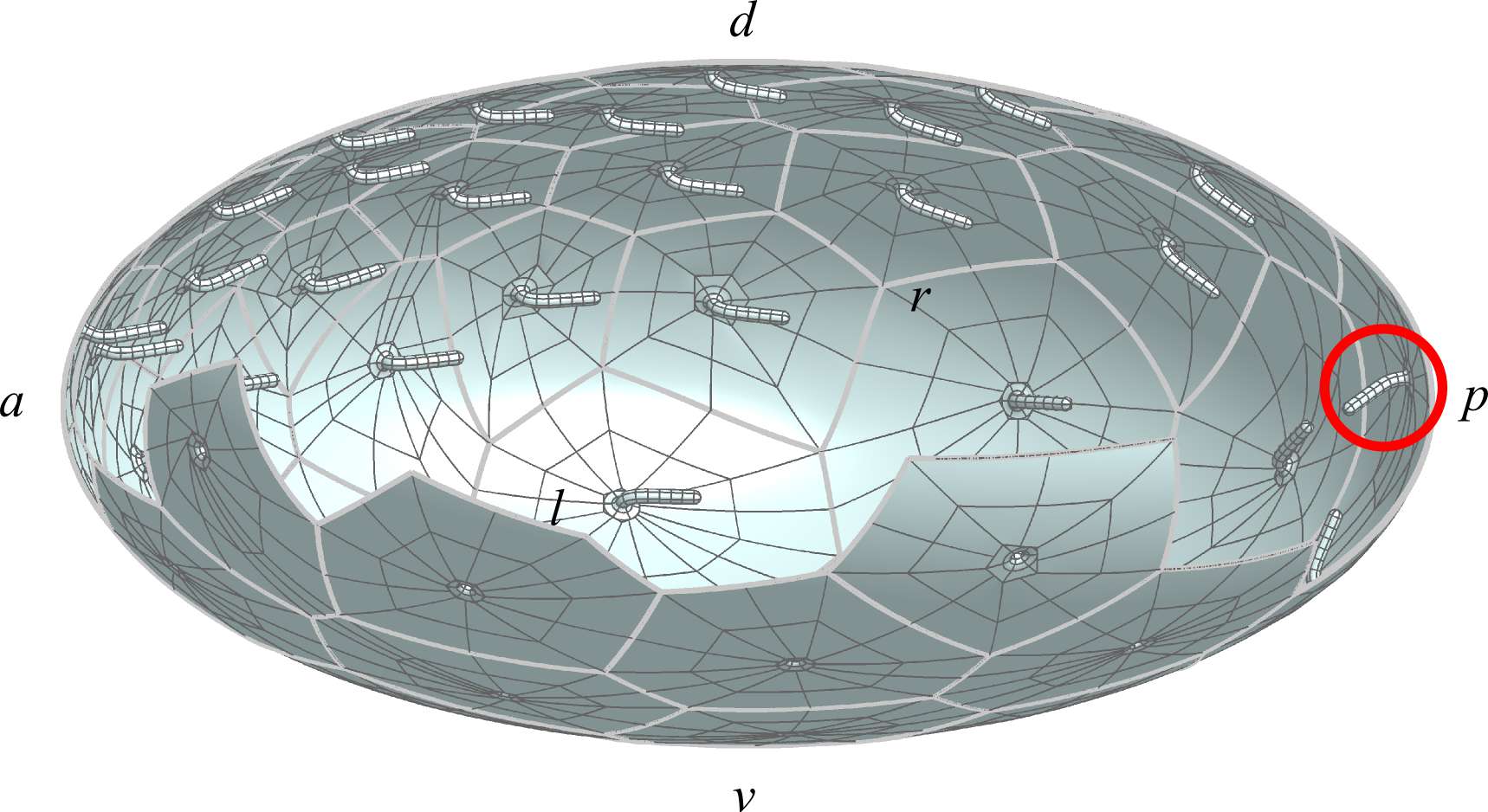}
	\caption{An example hybrid quadratic mesh of Kupffer's vesicle showing non-uniform distribution of cilia, all tilted towards the posterior, corresponding to figure \ref{fig:results-expt} (a), (b) and (c). A section has been cut away for internal viewing and comparison with figure \ref{fig:KV}. The posterior pole, at which we prescribe the tilt angle as zero, is indicated with a red circle (see text for details).
}
	\label{fig:KVmesh}
\end{figure}

Formalised mathematically as the Hairy Ball Theorem, attempting to specify a constant magnitude direction vector over the entire surface of the ellipsoid inevitably results in a point of discontinuity. For this reason we prescribe the tilt angle to reduce smoothly to zero at the poles where this occurs, therefore we have zero tilt at the posterior pole in the posterior tilt model (figure~\ref{fig:KVmesh}, indicated with a red circle) and zero tilt at the dorsal pole in the dorsal tilt model.

%Specifying a consistent tilt direction inevitably leads to regions where the tilt direction changes rapidly over a very small area, a consequence of the Hairy Ball Theorem. To rectify this, we reduce the tilt angle smoothly in these regions (figure~\ref{fig:KVmesh}), so that no two neighbouring cilia have greatly differing tilt (figure \ref{fig:tilt_dirn}). Combining these features results in a mesh of KV including all cilia for any given time (figure \ref{fig:KVmesh}).
%In order to avoid the inevitable cowlicks of the hairy ball theorem, we reduce the tilt angle in regions where the tilt direction changes quickly over a small region. We generate an unstructured mesh of quadratic triangles and quadrilaterals of the geometry.

\subsection{Numerical implementation}
\label{subsec:num-imp}

The mathematical problem to be solved is the determination of the unknown stress $f_i(\bm{y})$ for $\bm{y}\in D$ from the prescribed surface velocity $u_j(\bm{y})$. This is achieved by discretising the stress, and applying collocation. The stress is discretised as taking piecewise constant values $f_i[1],\ldots ,f_i[N]$ on surface elements $D[1],\ldots ,D[N]$, where $D=D[1]\cup \ldots \cup D[N]$ is the surface mesh and $N$ is the number of mesh elements.
The discrete form of equation \eqref{eq:rsBIE} is then given by,
\begin{equation}
u_j(\bm{y}) = \sum_
{n = 1}^{N}f_i[n]\int_{D[n]}{S_{ij}^\epsilon(\bm{x},\bm{y})\mathrm{d}S_{\bm{x}}} \quad \mbox{where} \quad \bm{y} \in D[m].
\label{eq:disc-sys}
\end{equation}

Taking $\bm{y}$ as the centroid of element $D[m]$ while allowing $m$ to range over $1,\ldots , N$ and $j=1,2,3$, we then have $3N$ equations for $3N$ unknown scalar stress variables. The numerical solution is carried out as described in the appendix.  Once the discrete approximations, $f_i[n]$, are calculated, the velocity field at any position in the domain can be found by reapplying equation~\eqref{eq:disc-sys}. This boundary element implementation of the regularized Stokeslet method, with the variable $f_{i}(\bm{x})$ being discretised separately from the numerical quadrature used to evaluate the regularized Stokeslet integrals, was shown by \citet{Smith09b} to give significant advantages in efficiency, an important requirement for solving problems with complex geometry. Note that the calculation described above applies to a specific instant in time; time-dependence was kept implicit in the explanation. In order to calculate the time-averaged flow over a full beat cycle, as in the results reported in section~\ref{sec:results}, the calculation must be repeated for a sequence of $N_t$ timesteps, $t=0, T/N_t, \ldots , T/(N_t-1)$ over the beat period $T$, the flow field for each timestep being stored and then averaged.

%This is then solved using collocation by applying \eqref{eq:disc-sys} at $\bm{y} = \bm{y}^m$, for $m = 1,\ldots,N$, centroids of surface elements. If we were using linear elements then the integrals in \eqref{eq:disc-sys} for $m = n$, the so-called diagonal entries of our system, would evaluate to zero exactly and would not contribute.

\section{Results}
\label{sec:results}

For comparison with the experimental results of \citet{Supatto08} shown in figure~\ref{fig:observed_flow}, we shall examine flow in a centrally located anterior-posterior left-right (AP-LR) plane ($x_1x_2-$plane in our model) at $x_3 = 0$. Results will be shown viewed from the dorsal side, looking in the ventral direction. Additionally, we will compute flow in transverse ($x_2=0$) sections; we are not aware of comparable experimental data for zebrafish KV, but the results may serve as experimentally-testable predictions, and also may be compared to known flow patterns in the different but related system of medakafish KV.

There are a number of aspects of the flow that could be examined, including the oscillatory component of the flow and the advection of particles by the time-dependent flow. In the present study we shall restrict our reports to the time-average of the instantaneous flow computed by the model over $60$ discrete intervals of the cilia beat cycle. Profiles will only be reported in the bulk of the fluid.

We examine how the flow field differs between models with:
\begin{enumerate}
\item [(1)] all cilia tilted towards the posterior pole \citep{Kramer05,Okabe08},
\item [(2)] all cilia tilted towards the dorsal pole \citep{Supatto08},
\item [(3)] mixed tilt directions as described in section \ref{subsec:mesh}.
\end{enumerate}
In each case, we compare flow with,
\begin{enumerate}
\item [(i)] a spatially homogeneous cilia distribution, and
\item [(ii)] a spatially varying cilia distribution with maximum density at the anterior of the dorsal roof \citep{Kreiling07}.
\end{enumerate}

% removed  -  from anterior to left in the above

%All results presented are viewed from dorsal, looking towards ventral unless otherwise stated.

\subsection{All cilia tilted posteriorly}
\label{subsec:post-tilt}

For homogeneous cilia density, with posterior tilt only, there is no evidence of an overall clockwise flow (figure~\ref{fig:results-even}a).
The beat direction is clockwise viewed from tip to base, however the cell surfaces of the dorsal roof and ventral floor face in opposite directions. Hence dorsal roof and ventral floor cilia perform opposite motions. Because cilia density is equal on both surfaces, the overall effect is a nearly zero flow in the AP-LR midplane. A posterior tilt makes no difference to this; flow components in the AP-LR midplane are still equal and opposite. As in all results shown, the largest magnitude flow occurs close to the walls of the vesicle since this is where the cilia are located.

By comparison, inhomogeneous cilia density does produce an overall anticlockwise flow viewed from dorsal (figure~\ref{fig:results-expt}a) because the opposing rotation directions are no longer equal; the flow produced by the dorsal roof cilia, rotating clockwise when viewed from dorsal, base to tip, predominates over that produced by the ventral floor cilia. The magnitude of the flow is however generally significantly less than the typical $10\mu\mathrm{m}/\mathrm{s}$ reported by \citet{Supatto11}. Our simulation results for this configuration show that the centre of the whirl in the $x_3=0$ plane is moved towards the anterior end of KV.

%specification of the geometry.

\subsection{All cilia tilted dorsally}
\label{subsec:dorsal-tilt}

%By contrast, a vesicle with exclusively dorsally tilted cilia produces anticlockwise flow in the AP-LR midplane for both homogeneous (figure~\ref{fig:results-even}d) and inhomogeneous (figure~\ref{fig:results-expt}d) cilia distributions.

The cilia motion in our simulations is always clockwise about its axis of rotation, viewed from tip to base. However, examining the cilia motion about the DV ($x_3$) axis, a dorsally-tilted cilium located around the equatorial region ($x_3=0$) will spend half of its beat cycle moving with an anticlockwise component relative to the DV axis, and half of its beat cycle moving with an clockwise component relative to the DV axis. The dorsal tilt results in equatorial cilia consistently performing an effective stroke in the anticlockwise direction relative to the DV axis,  and a recovery stroke in the clockwise direction relative to the DV axis. The anticlockwise flow relative to the DV axis will predominate, as evident in the anticlockwise pointing arrows in figure~\ref{fig:results-even}d (homogeneous distribution), and figure~\ref{fig:results-expt}d (higher concentration on the dorsal roof). Comparing figures~\ref{fig:results-even}d and \ref{fig:results-expt}d, it is evident that with cilia concentrated on the anterior of the dorsal roof, the centre of the whirl is again moved to the anterior, moreover the rotational flow velocity is increased relative to the homogeneous case.

\subsection{Cilia tilted dorsally and posteriorly}
\label{subsec:both-tilt}

\begin{figure}
	\centering
\[
\begin{array}{c}
\begin{array}{ll}
    \begin{array}{c}
        \mbox{(a) Posterior tilt} \\
        \raisebox{0.12cm}{
            \includegraphics[width=6.2cm]{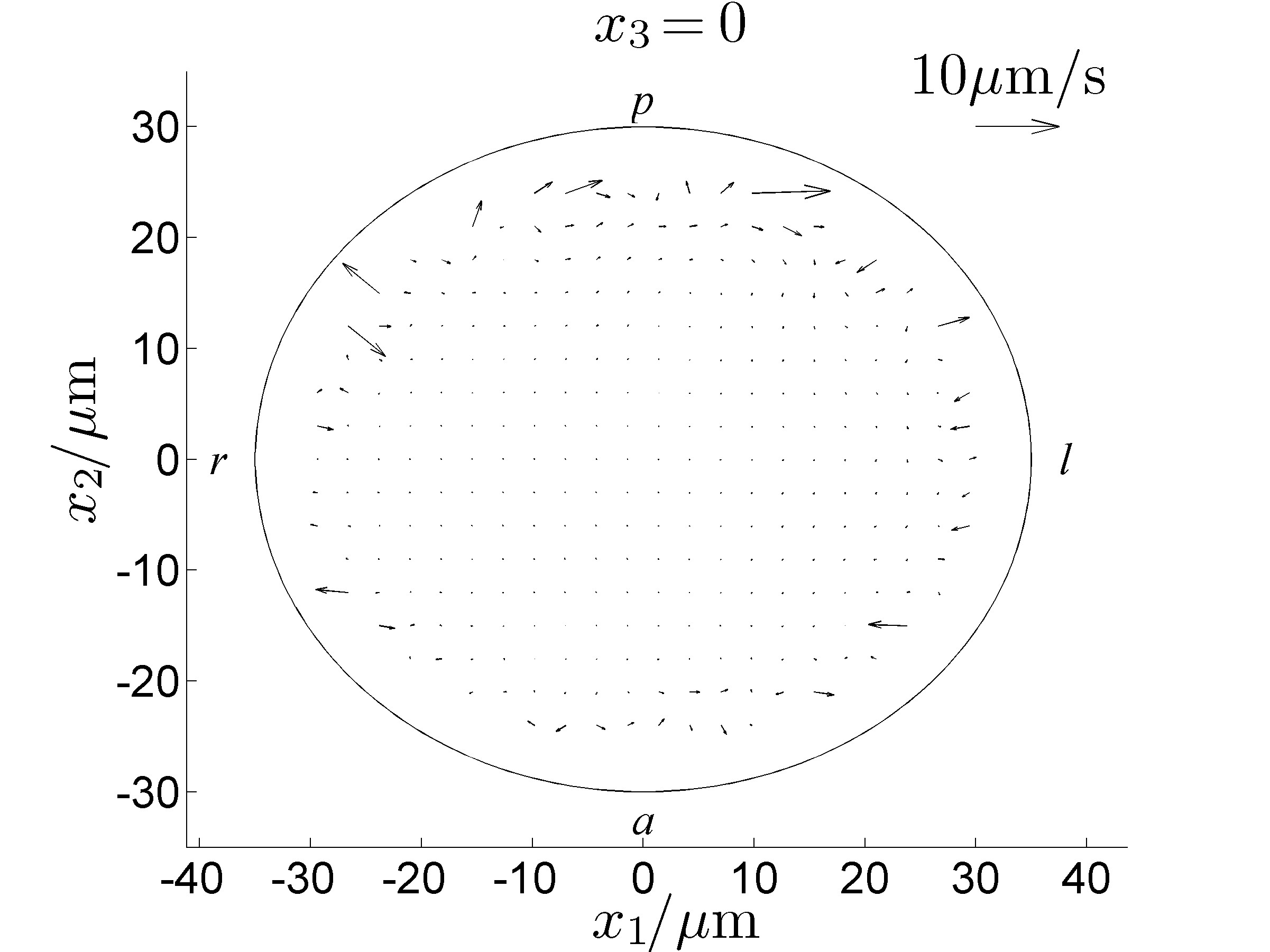}
        }
    \end{array}
    &
    \begin{array}{l}
        \mbox{(b)} \\
        \includegraphics[width=3.3cm]{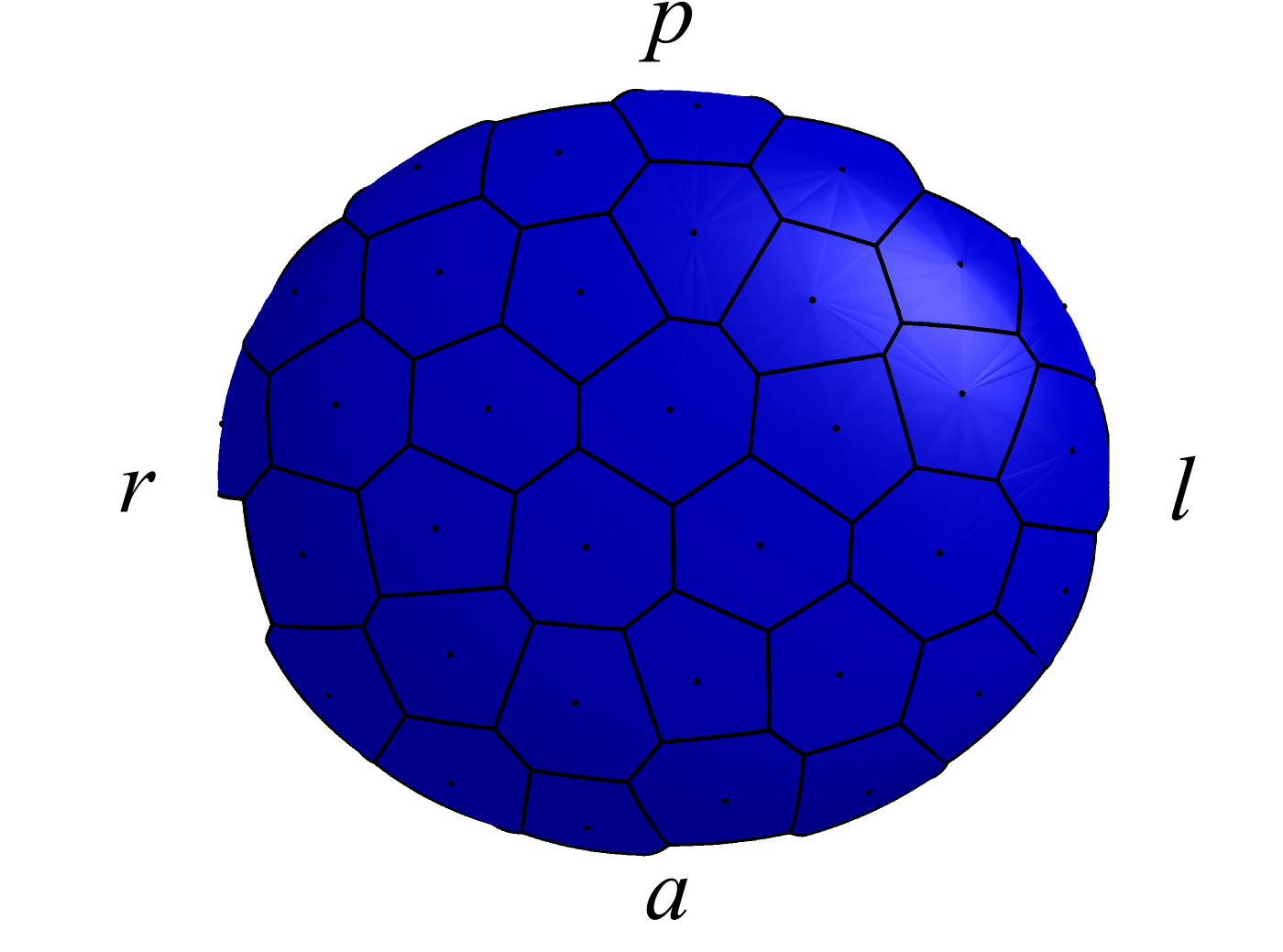} \\
        \mbox{(c)} \\
        \includegraphics[width=3.3cm]{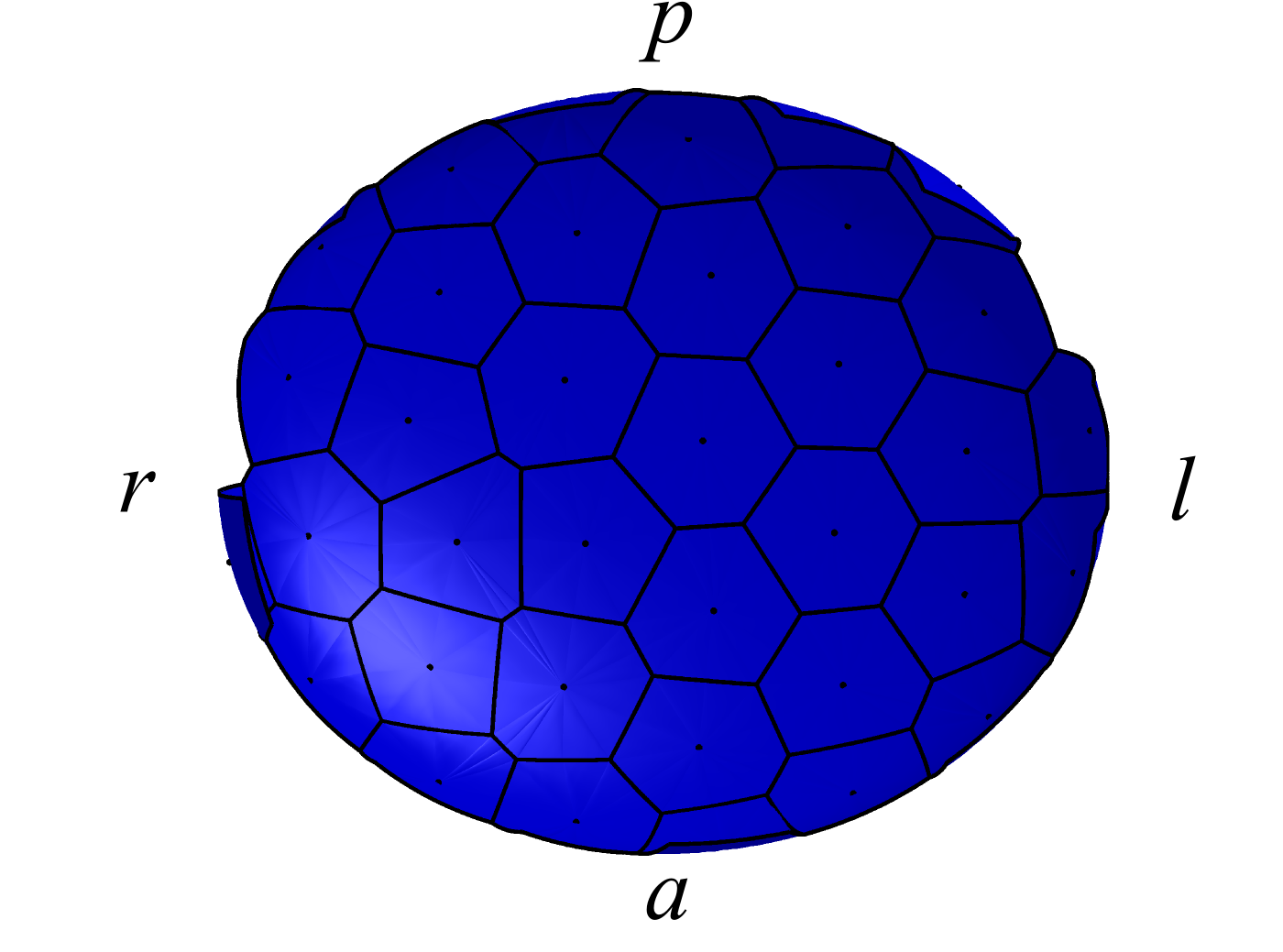}
    \end{array}
\end{array}
\\
\begin{array}{ll}
    \begin{array}{c}
        \mbox{(d) Dorsal tilt} \\
        \raisebox{0.12cm}{
            \includegraphics[width=6.2cm]{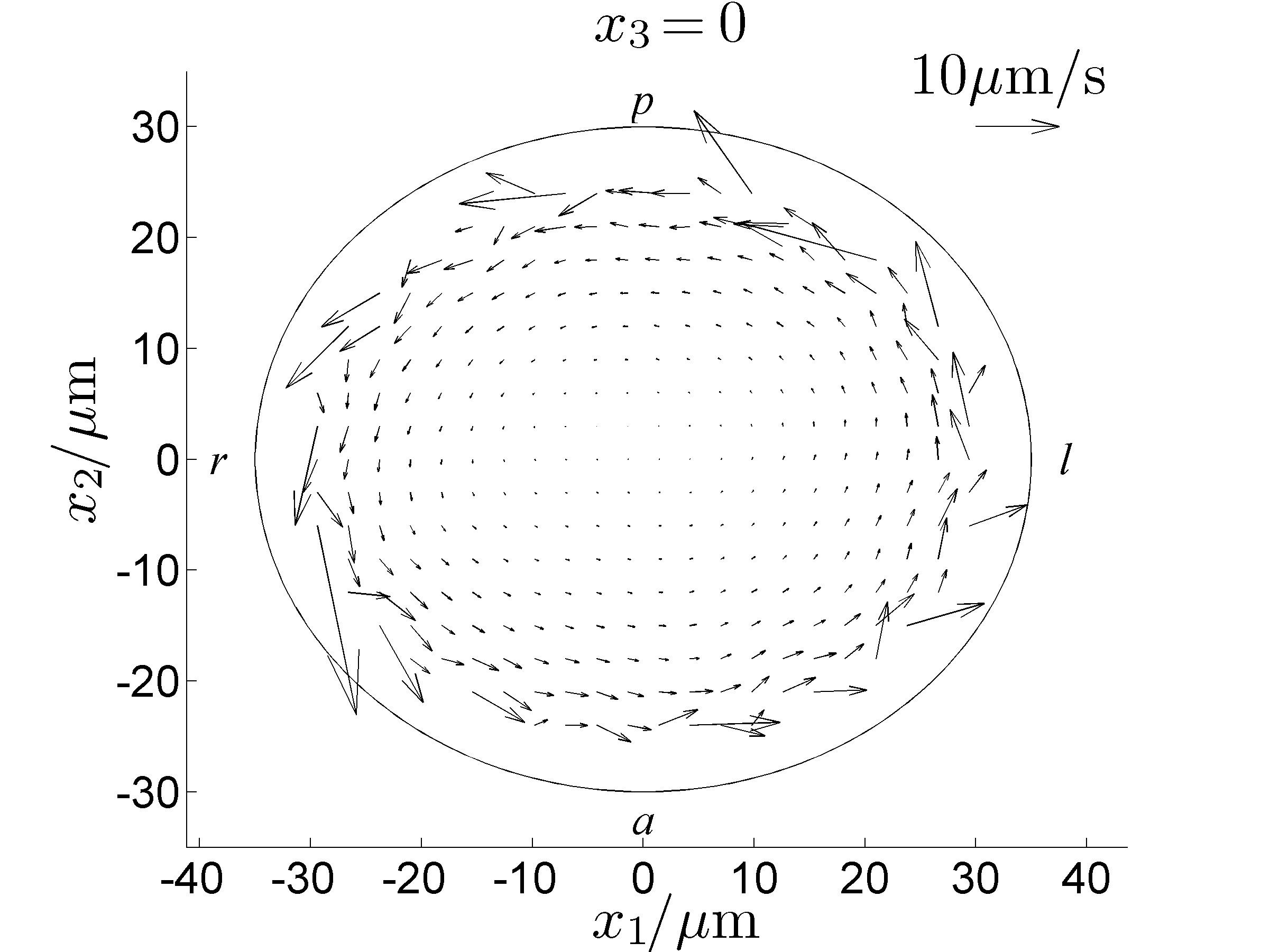}
        }
    \end{array}
    &
    \begin{array}{l}
        \mbox{(e)} \\
        \includegraphics[width=3.3cm]{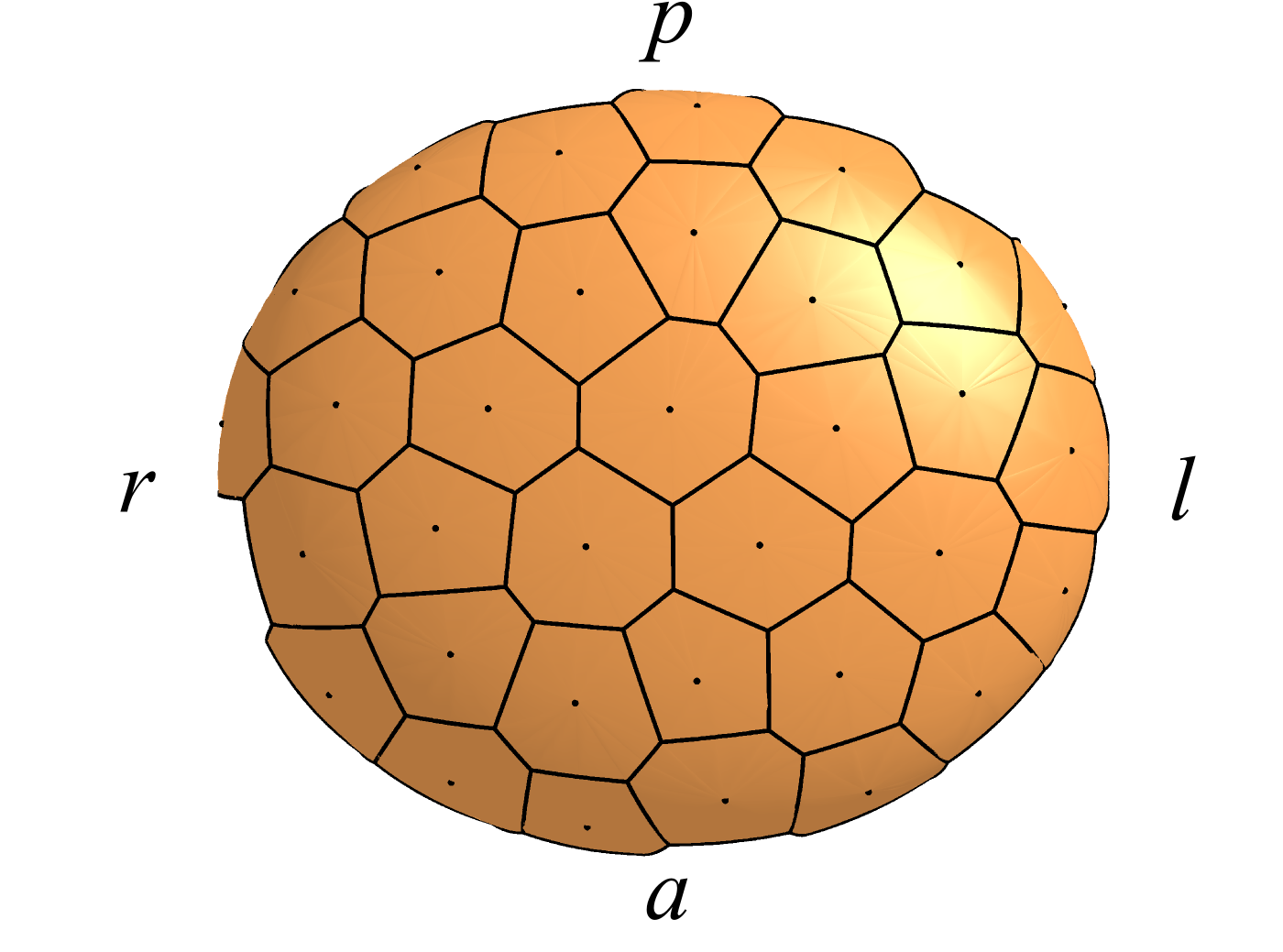} \\
        \mbox{(f)} \\
        \includegraphics[width=3.3cm]{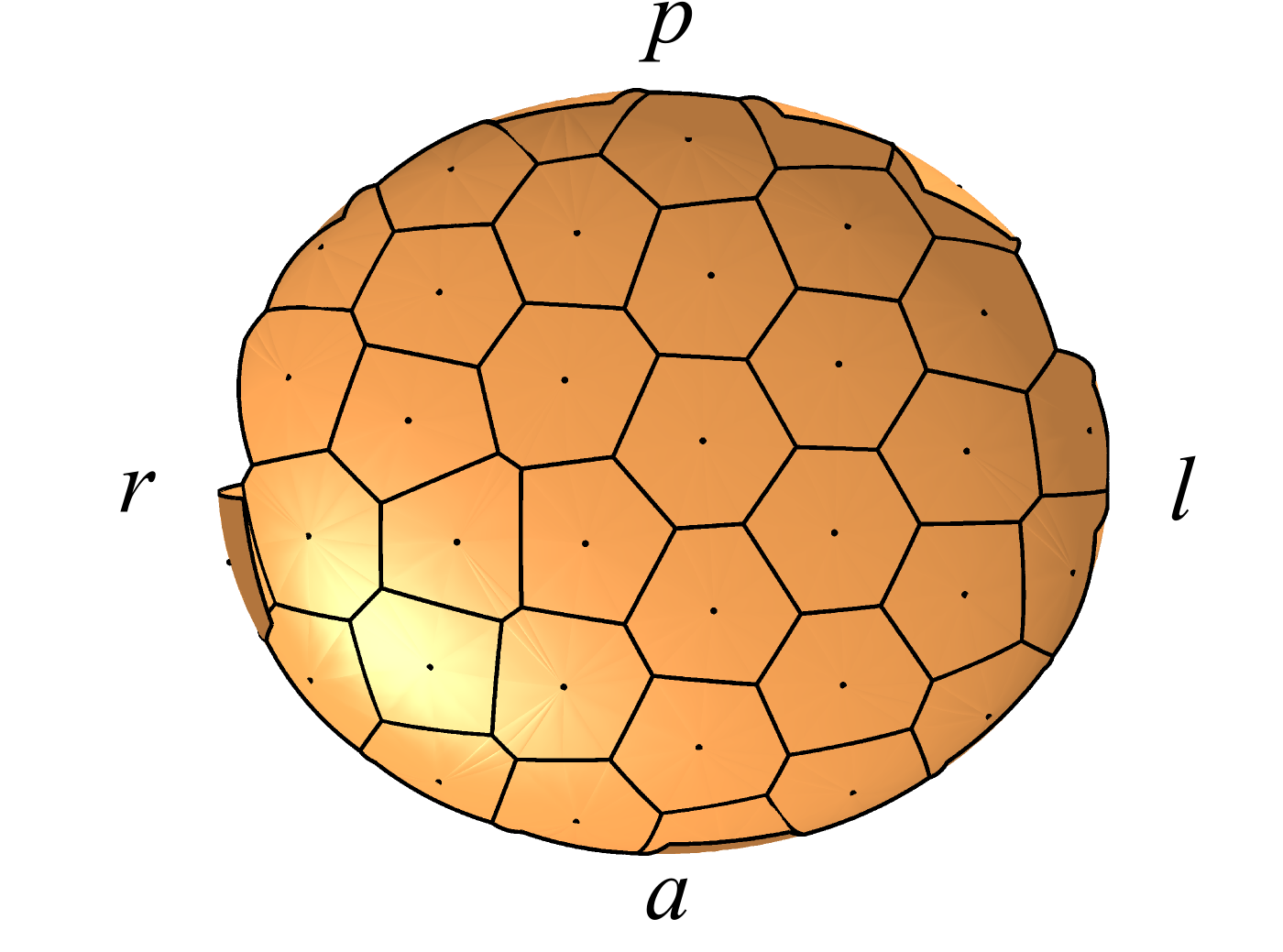}
    \end{array}
\end{array}
\\
\begin{array}{ll}
    \begin{array}{c}
        \mbox{(g) Mixed posterior and dorsal tilt} \\
        \raisebox{0.12cm}{
            \includegraphics[width=6.2cm]{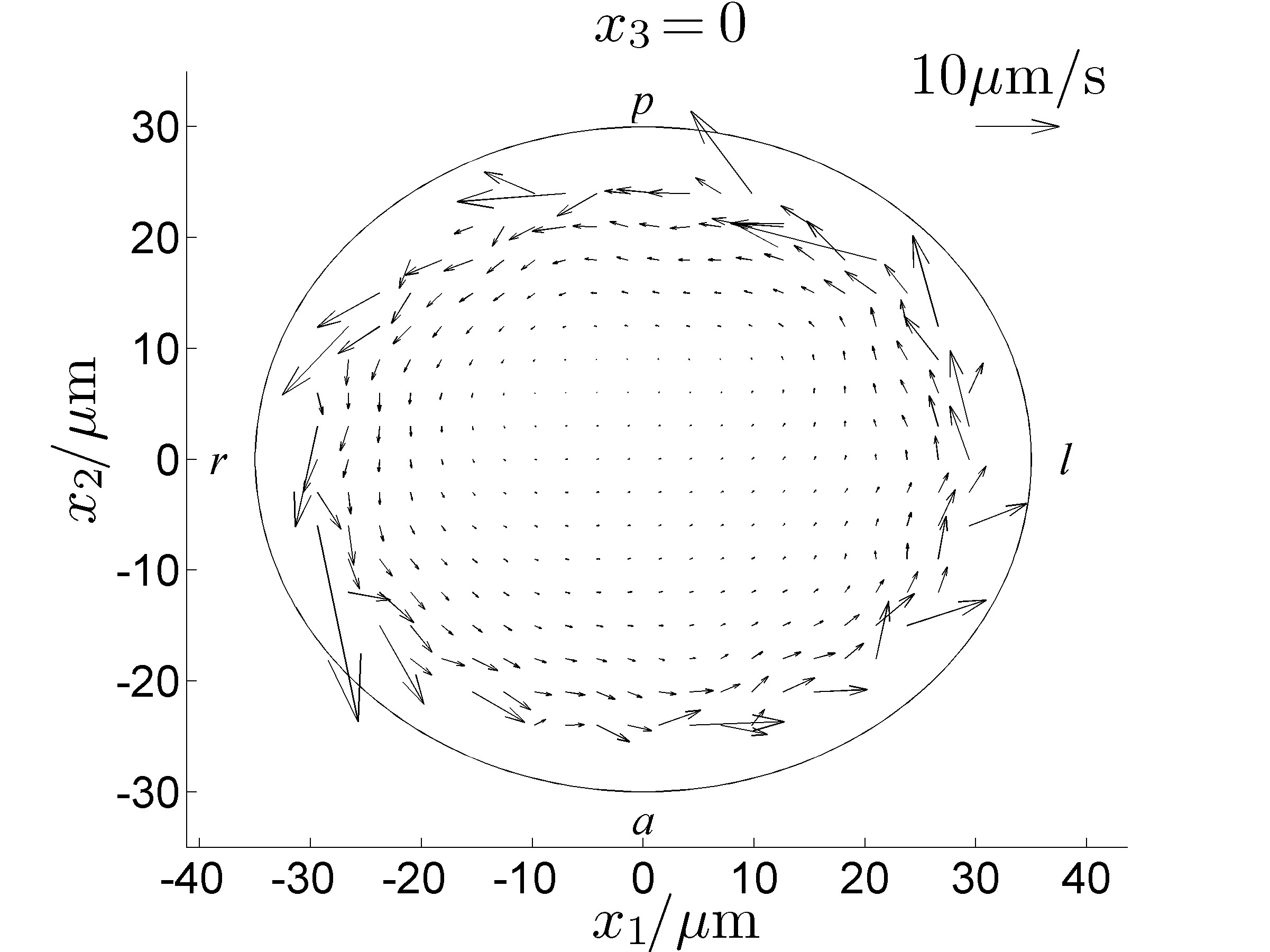}
        }
    \end{array}
    &
    \begin{array}{l}
        \mbox{(h)} \\
        \includegraphics[width=3.3cm]{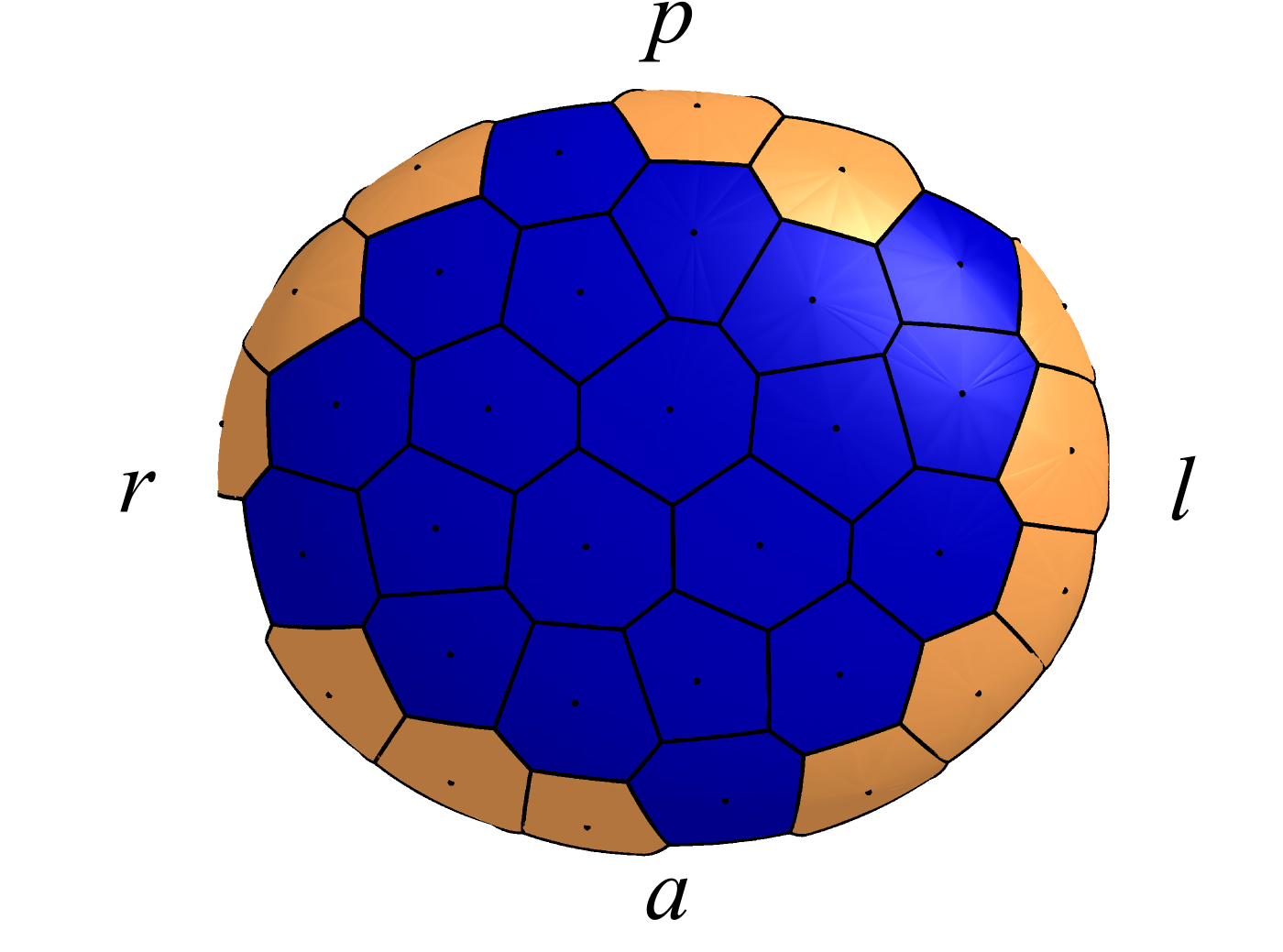} \\
        \mbox{(i)} \\
        \includegraphics[width=3.3cm]{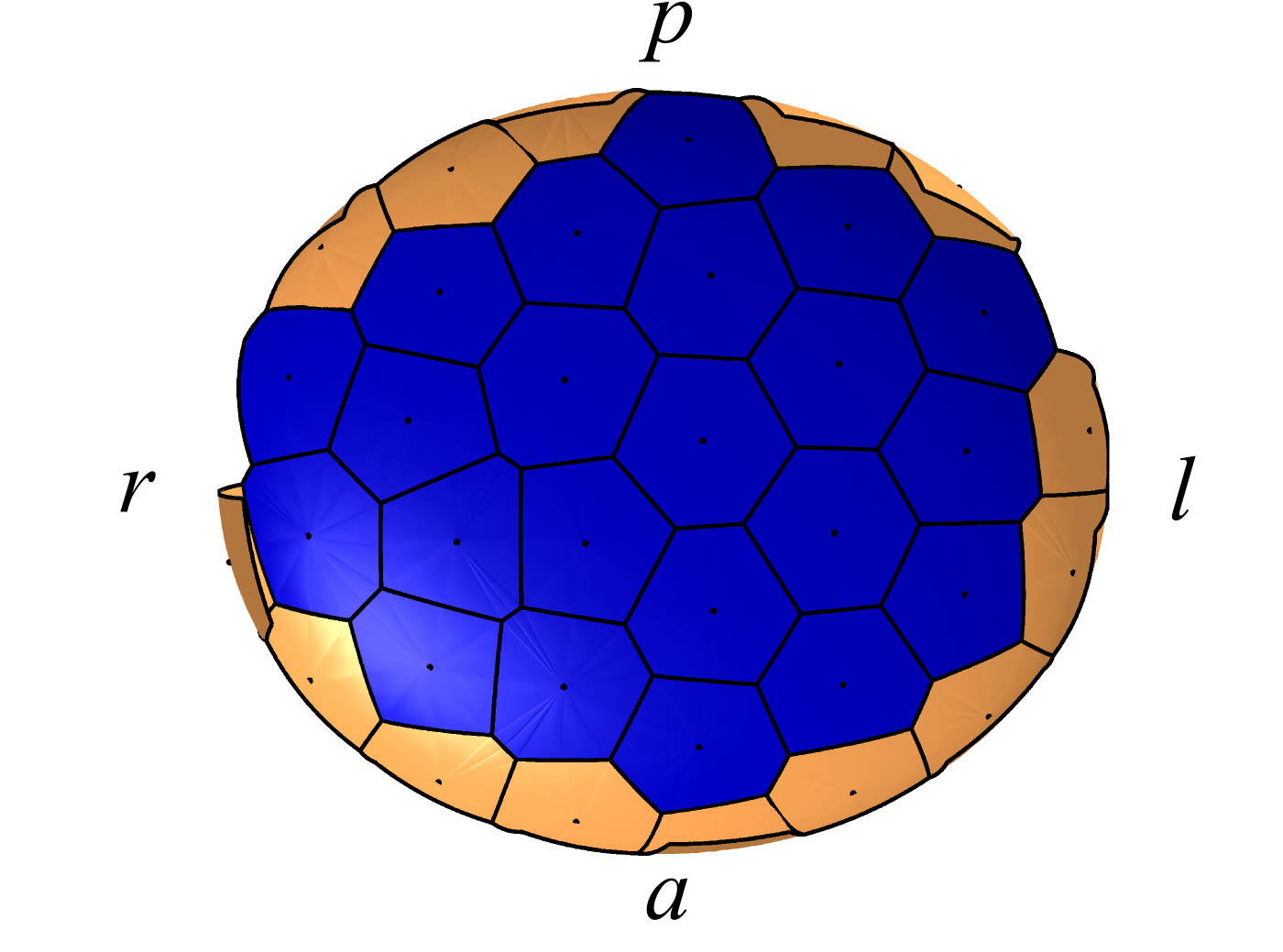}
    \end{array}
\end{array}
\end{array}
\]
	\caption{Time-averaged flow fields $(u_1(\bm{x}),u_2(\bm{x}))$ in the AP-LR plane $x_3 = 0$ computed over a beat cycle, looking from dorsal towards ventral, for three different specifications of cilia tilt, posterior, dorsal and mixed. Results were computed with a spatially homogeneous distribution of cilia. Left hand panels (a,d,g) show flow fields, right hand panels show the cell, and hence cilia, density on the dorsal roof (b,e,h) and ventral floor (c,f,i); red denotes posterior tilt, blue denotes dorsal tilt. For the mixed tilt case (h,i), equatorial cilia are dorsally-tilted, the remainder are posteriorly tilted. Axis notation: {\it a} anterior; {\it p}, posterior; {\it l}, left; {\it r}, right. Arrow scale of $10\; \mu\mathrm{m}/\mathrm{s}$ corresponds to a prescribed cilia beat frequency of $\omega=30$ Hz and cilium length $L=3\; \mu$m; due to the linearity of Stokes flow, results for different values of cilium length or frequency can be inferred from the relationship $u\propto \omega L$. Only time-averaged flow above the cilia tips is shown.
}
	\label{fig:results-even}
\end{figure}

\begin{figure}
	\centering\[
\begin{array}{c}
\begin{array}{ll}
    \begin{array}{c}
        \mbox{(a) Posterior tilt} \\
        \raisebox{0.12cm}{
            \includegraphics[width=6.2cm]{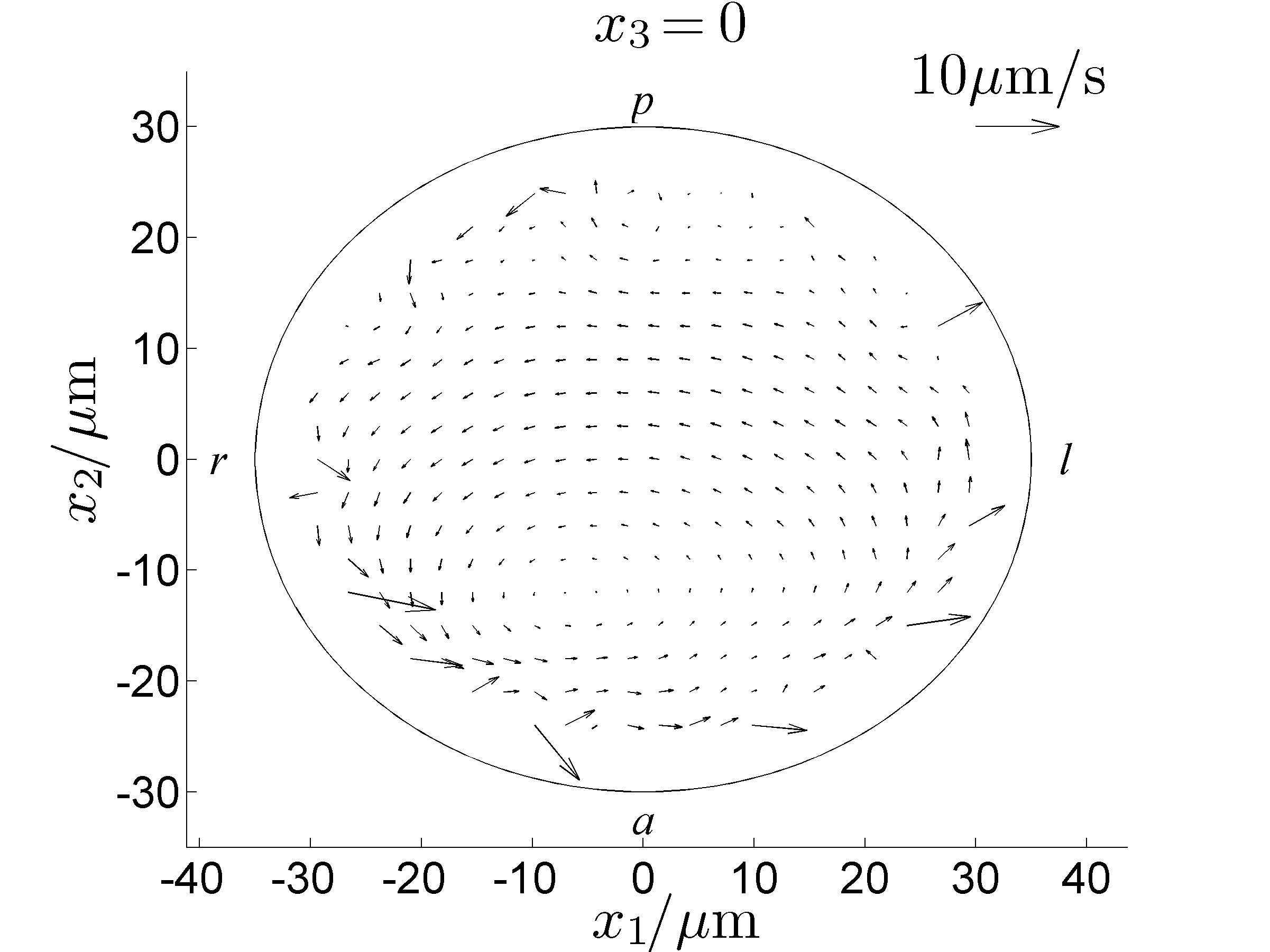}
        }
    \end{array}
    &
    \begin{array}{l}
        \mbox{(b)} \\
        \includegraphics[width=3.3cm]{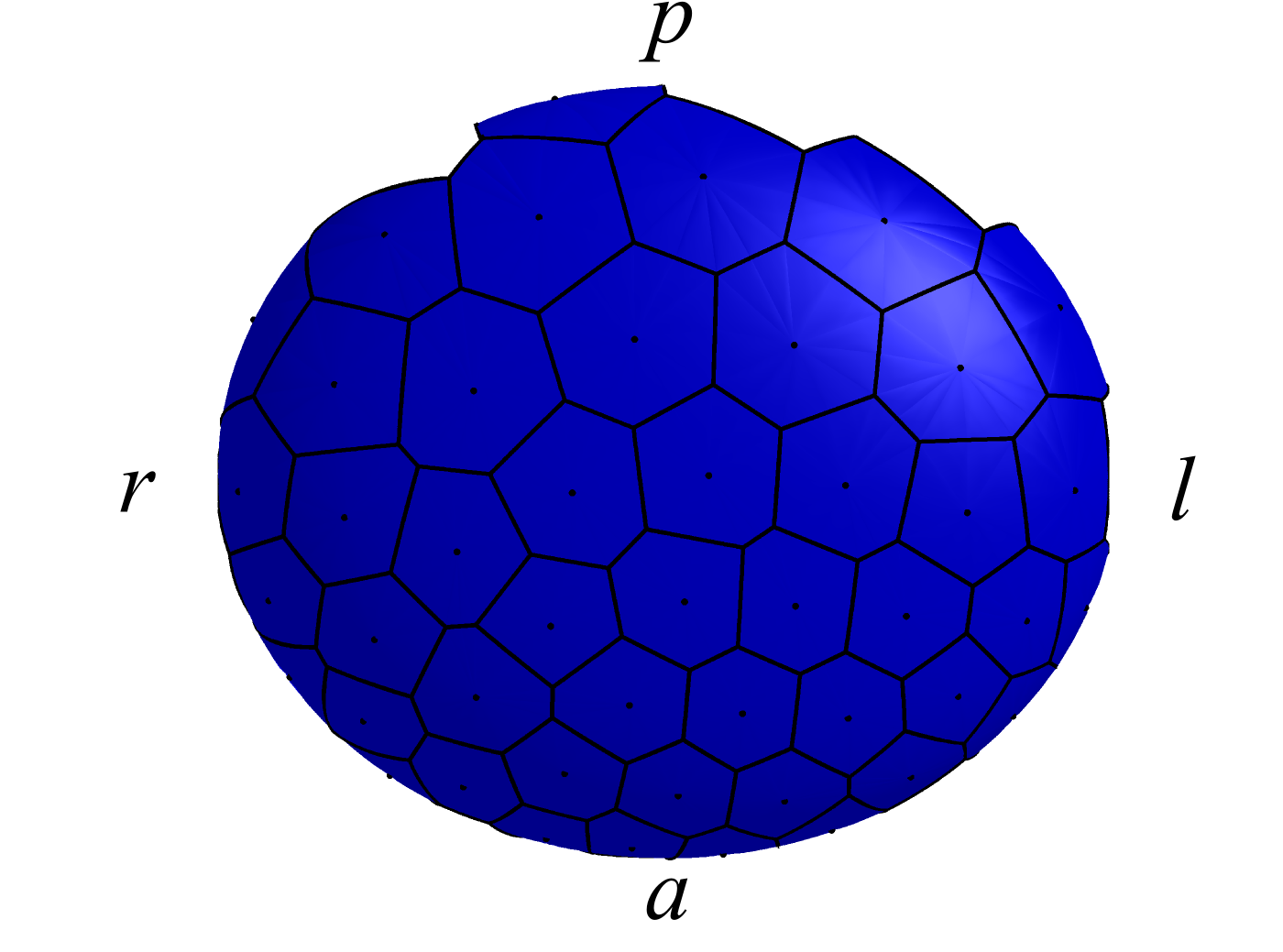} \\
        \mbox{(c)} \\
        \includegraphics[width=3.3cm]{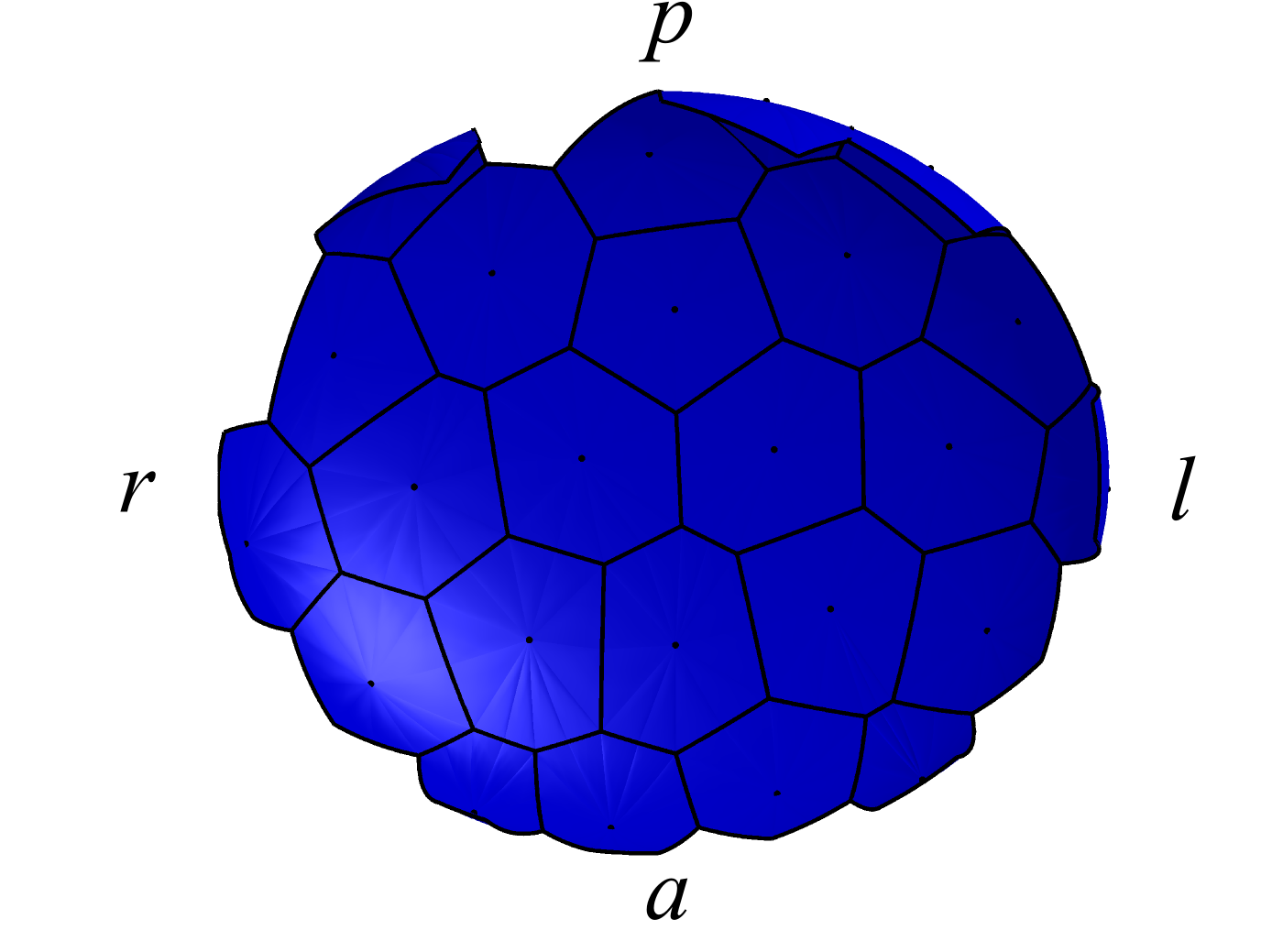}
    \end{array}
\end{array}
\\
\begin{array}{ll}
    \begin{array}{c}
        \mbox{(d) Dorsal tilt} \\
        \raisebox{0.12cm}{
            \includegraphics[width=6.2cm]{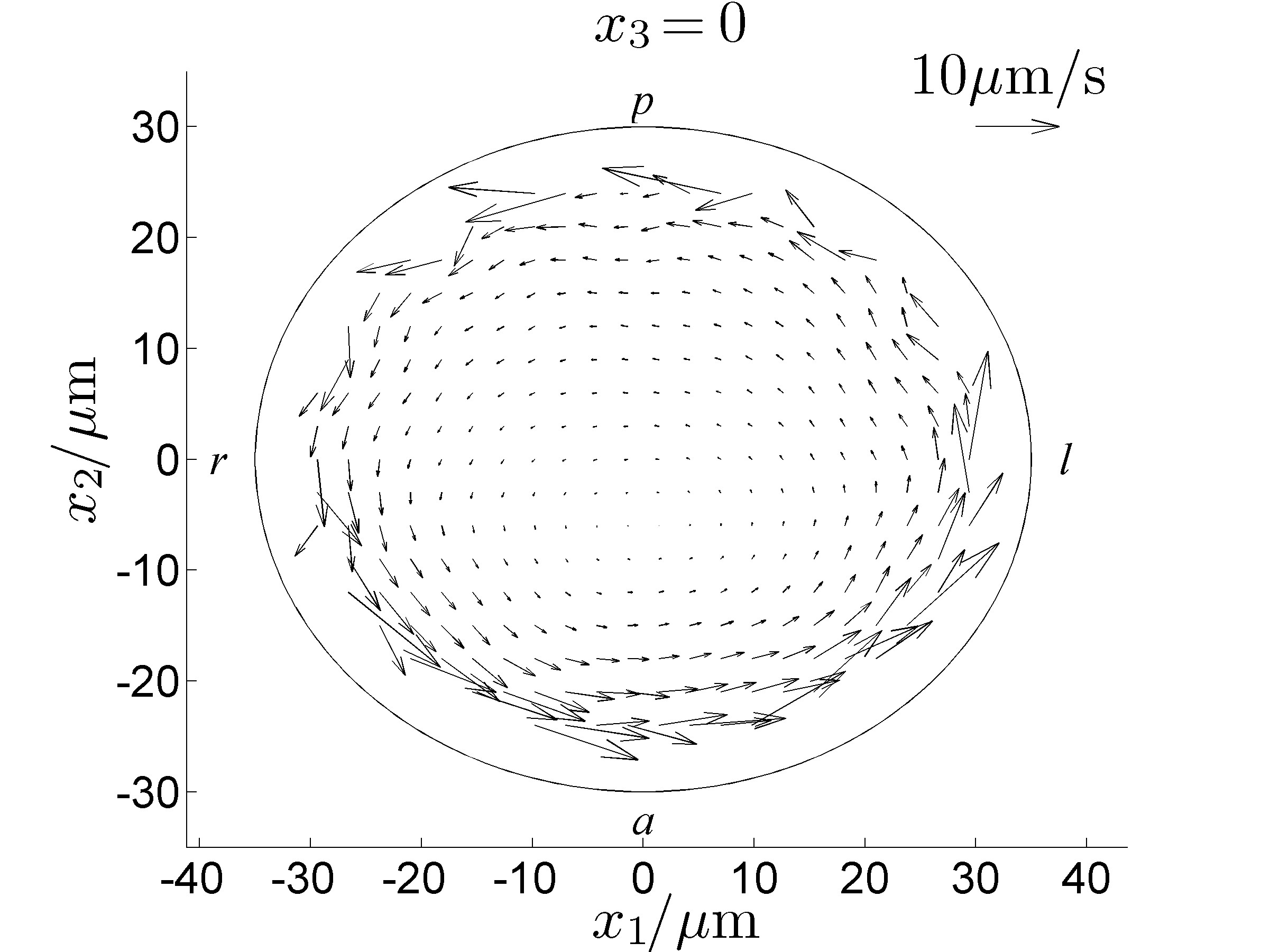}
        }
    \end{array}
    &
    \begin{array}{l}
        \mbox{(e)} \\
        \includegraphics[width=3.3cm]{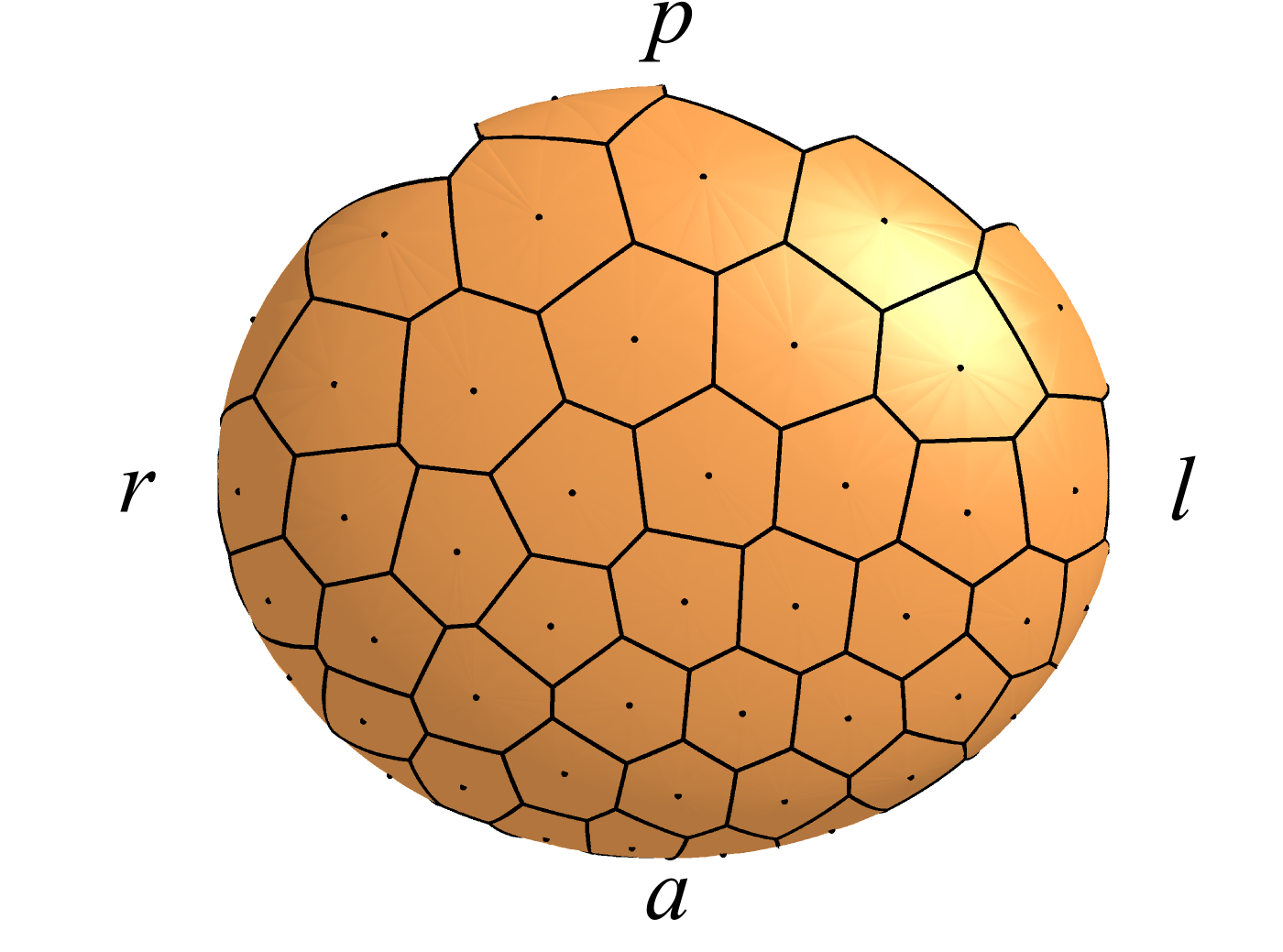} \\
        \mbox{(f)} \\
        \includegraphics[width=3.3cm]{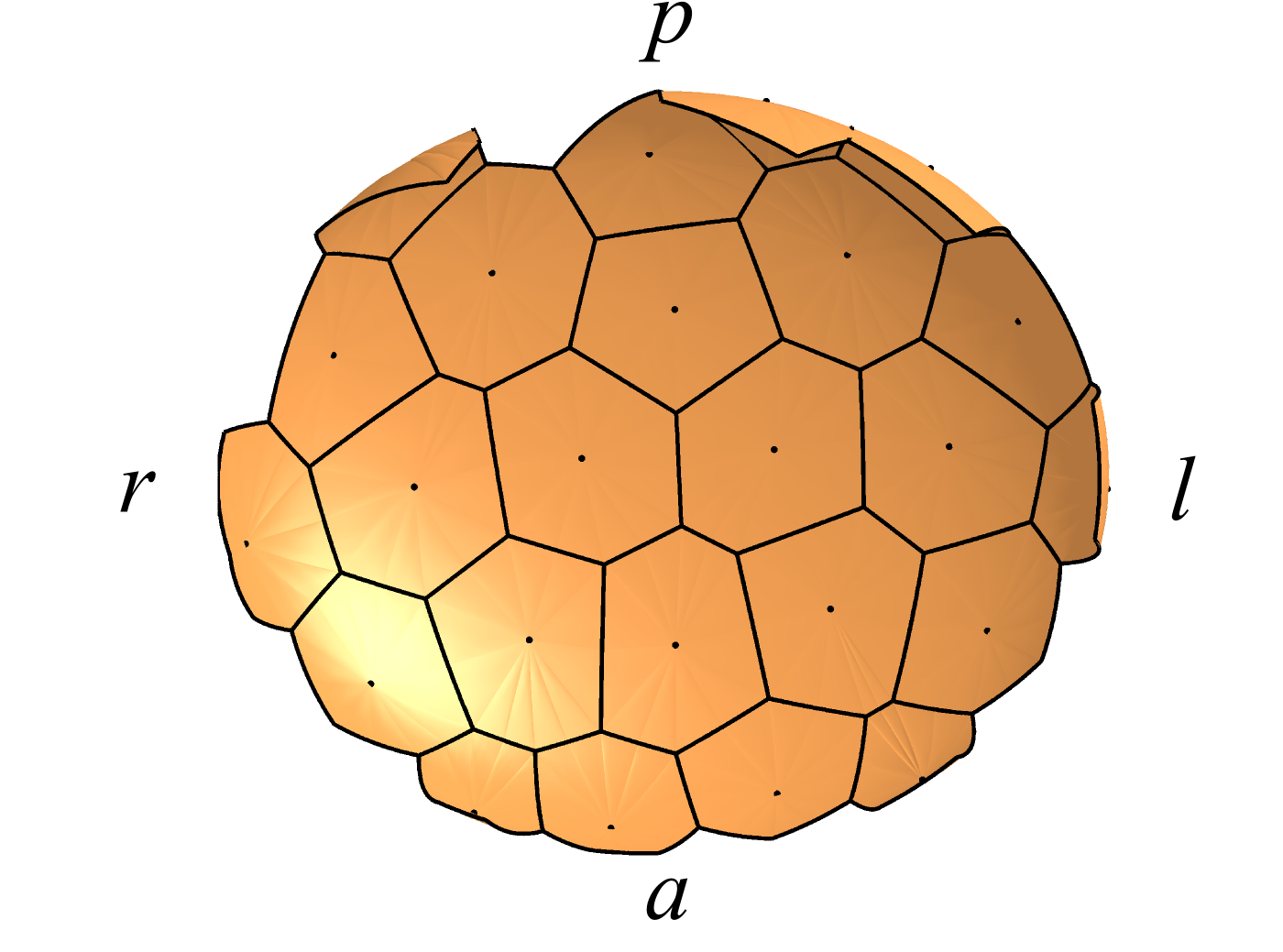}
    \end{array}
\end{array}
\\
\begin{array}{ll}
    \begin{array}{c}
        \mbox{(g) Mixed posterior and dorsal tilt} \\
        \raisebox{0.12cm}{
            \includegraphics[width=6.2cm]{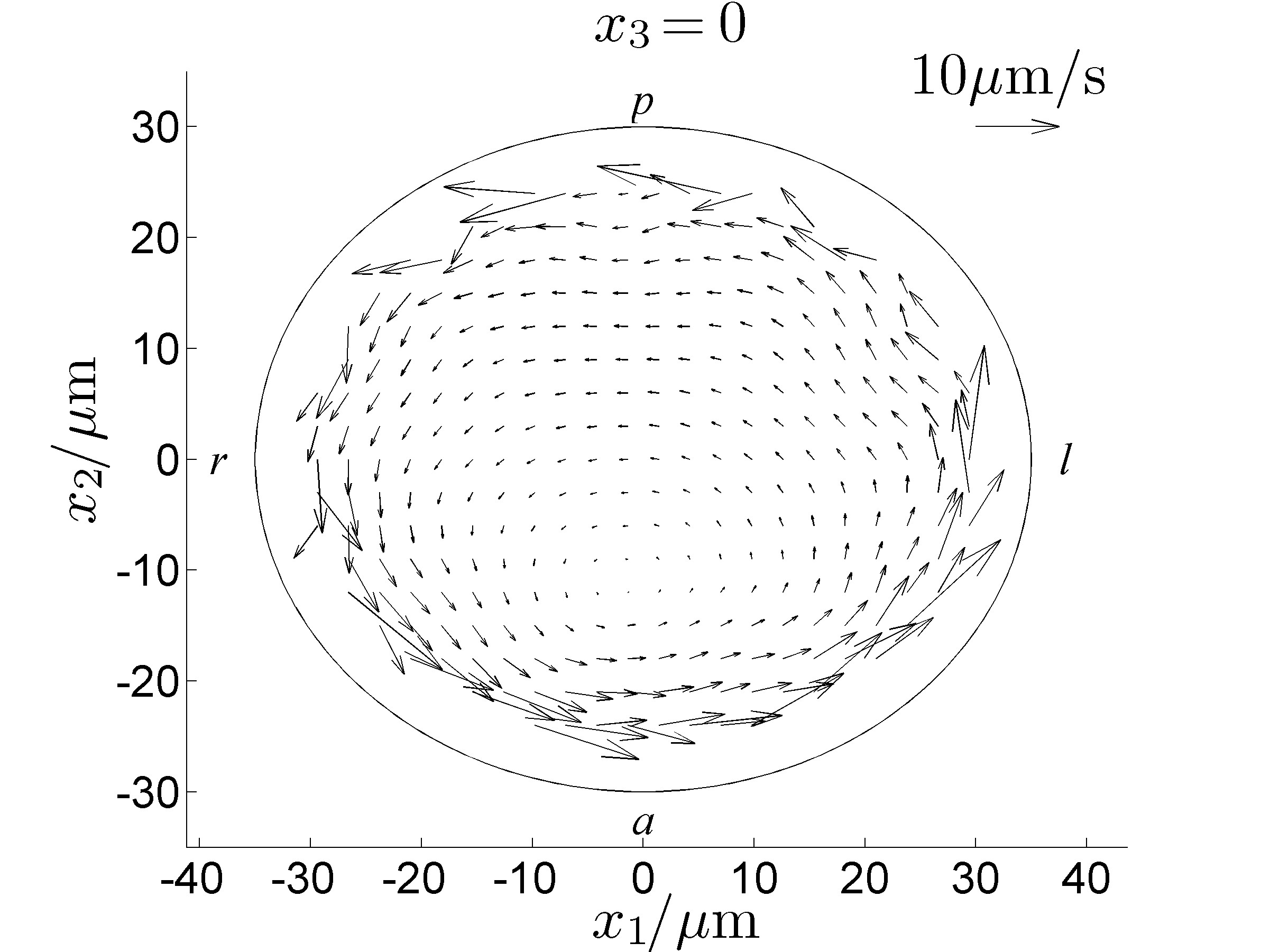}
        }
    \end{array}
    &
    \begin{array}{l}
        \mbox{(h)} \\
        \includegraphics[width=3.3cm]{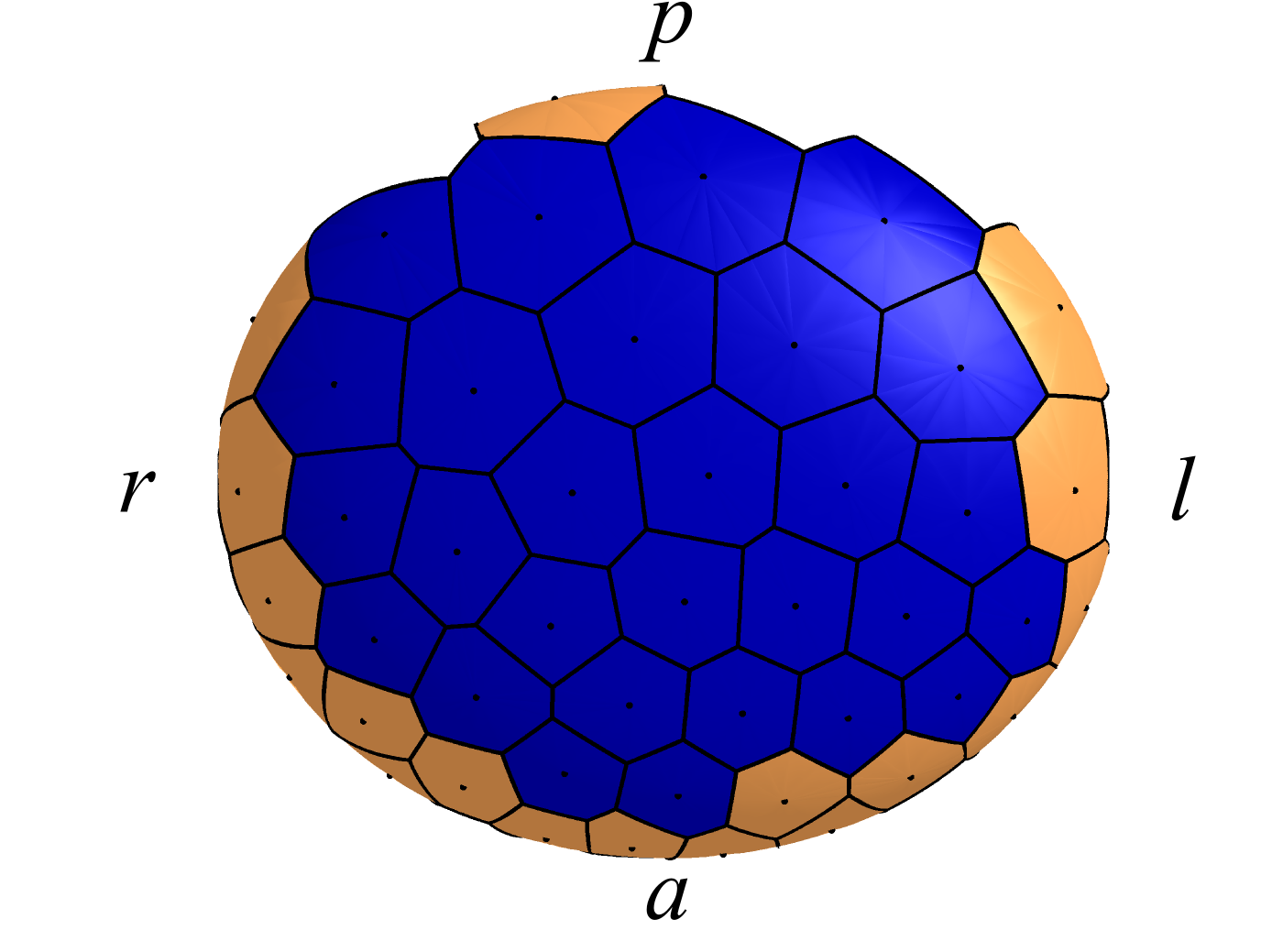} \\
        \mbox{(i)} \\
        \includegraphics[width=3.3cm]{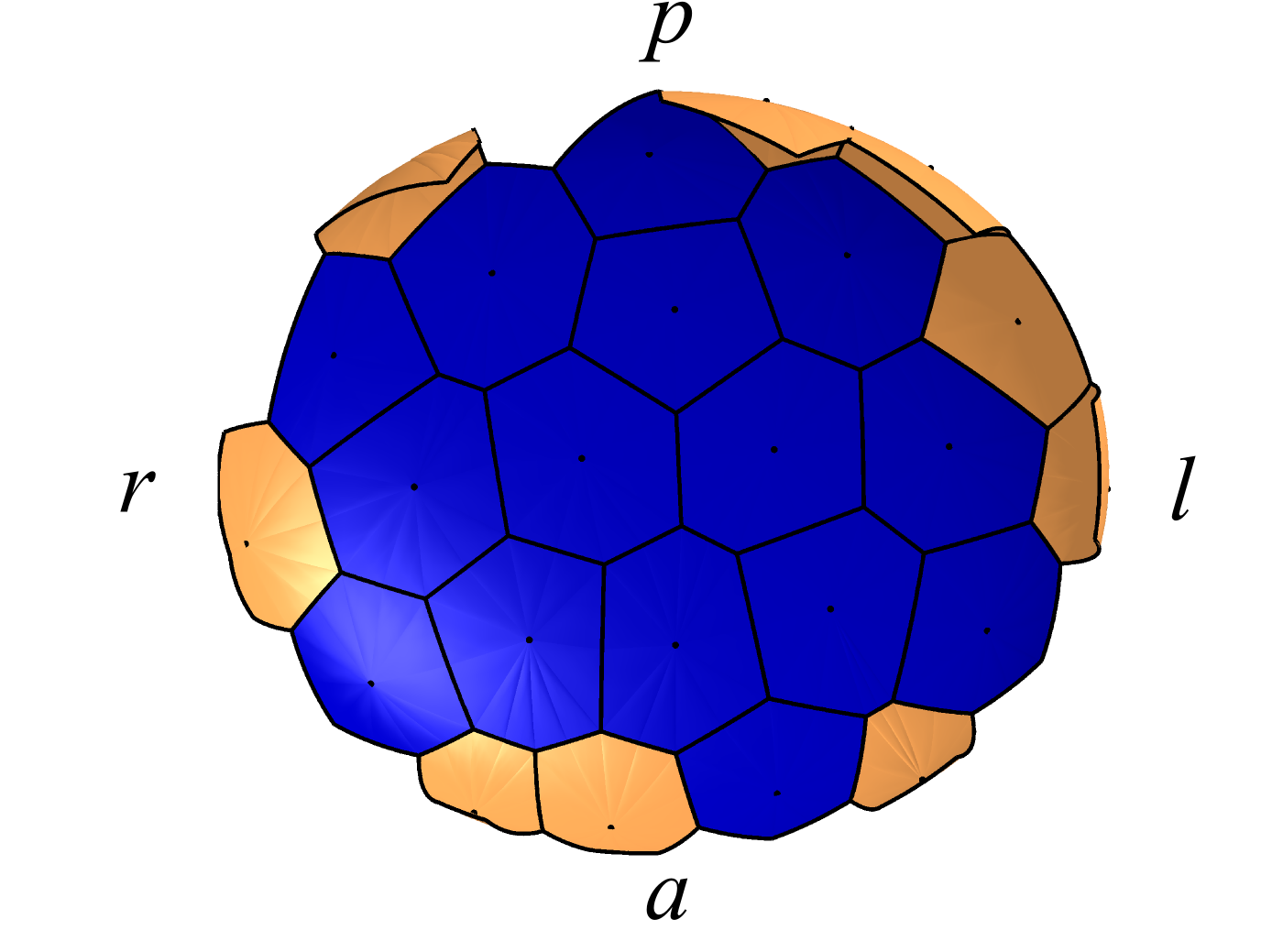}
    \end{array}
\end{array}
\end{array}
\]
	\caption{Time-averaged flow fields at $x_3 = 0$ for inhomogeneous cilia density distribution based on experimental observations, as shown in figure~\ref{fig:dorsal_ventral}. Other details are as shown in figure~\ref{fig:results-even}.}
	\label{fig:results-expt}
\end{figure}

Results for a mixture of posterior tilt on the roof and floor, and dorsal tilt at the equator, are shown in figures~\ref{fig:results-even}g and \ref{fig:results-expt}g. For homogeneous cilia distribution, the results are very similar to the pure dorsal tilt model, however with increased cilia density on the anterior of the dorsal roof, the mixed tilt results in increased flow magnitude.
This anticlockwise global vortex that is strongest from right to anterior to left is similar qualitatively to the experimental observations \citep{Kreiling07,Supatto08}; the flow is of similar magnitude in the region above the cilia tips, and has the largest magnitude of the models considered.

\begin{figure}
	\centering\[
\begin{array}{l}
    \begin{array}{c}
        \mbox{(a) Dorsal tilt} \\
        \raisebox{0.0cm}{
            \includegraphics[width=7.8cm,viewport=0in 3.125in 8.22in 7.5in,clip]{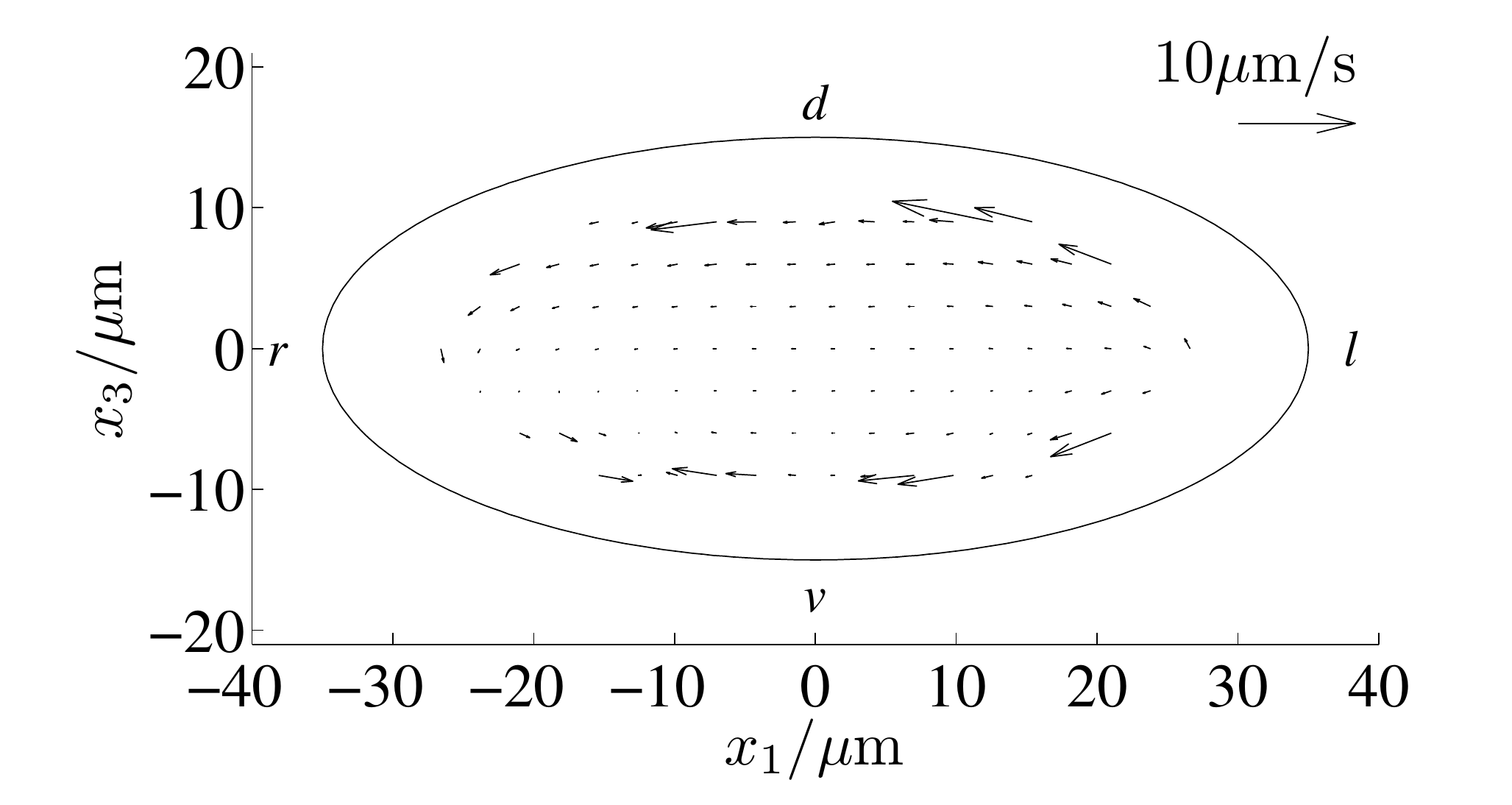}
        }
    \end{array}
    \\
    \begin{array}{c}
       \mbox{(b) Mixed posterior and dorsal tilt} \\
       \raisebox{0.0cm}{
           \includegraphics[width=7.8cm,viewport=0in 3.125in 8.22in 7.5in,clip]{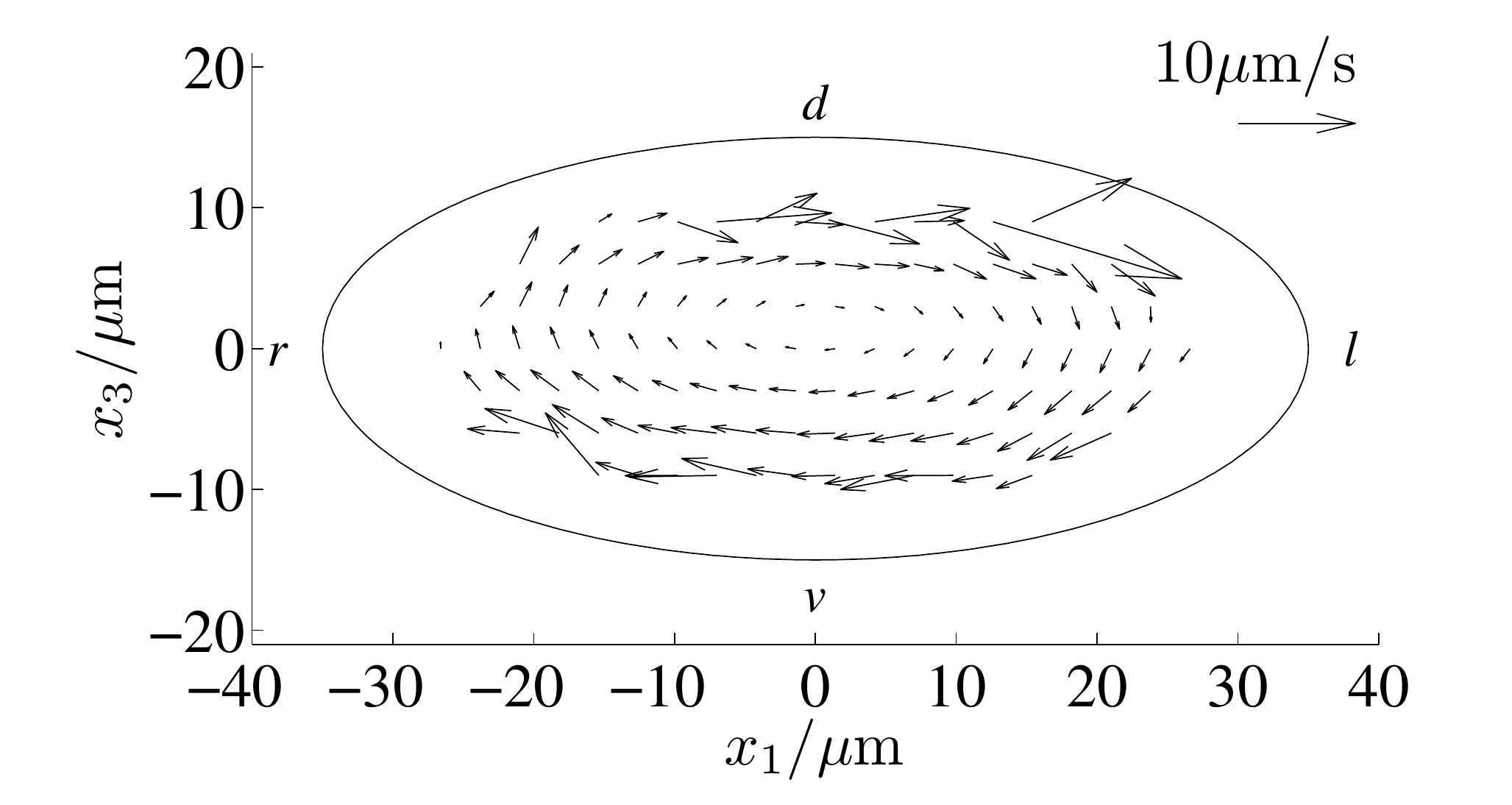}
       }
    \end{array}
\end{array}
\]
	\caption{Time-averaged flow fields at $x_2 = 0$ (transverse plane equidistant from anterior and posterior poles) for the inhomogeneous cilia density distribution based on experimental observations (figure~\ref{fig:dorsal_ventral}). Results for dorsal-only tilt are shown in (a), and for a mixture of dorsal and posterior tilt in (b). The latter shows a leftward flow near the dorsal roof, and rightward counterflow flow near the ventral floor. Other details are as shown in figure~\ref{fig:results-even}.}
	\label{fig:results-transverse}
\end{figure}

\subsection{Comparison of dorsal-only and mixed tilt}
While the equatorial sections shown in figures~\ref{fig:results-expt}(d,g) do not show great differences between dorsal-only and mixed posterior and dorsal tilt, analysis of flow in a transverse section equidistant from the anterior and posterior poles, shows considerable differences. The flow field resulting from exclusively dorsally-tilted cilia is relatively small and does not show any clear directionality (figure~\ref{fig:results-transverse}a). By contrast, a mixture of posterior and dorsal tilt (figure~\ref{fig:results-transverse}b) results in a leftward flow near the dorsal roof, and rightward return flow near the ventral floor of similar magnitude.
%This is similar to the combination of leftward flow and counterflow reported in the different system of medakafish KV. As may be expected, the additional layer of cilia on the ventral floor in zebrafish \cite{Okabe08} appears to result in an augmented rightward flow velocity relative to that observed in medakafish \cite{Okada05}.

\section{Summary and discussion}
\label{sec:summary}

Previous experimental studies of cilia-driven flow inside Kupffer's vesicle (KV) established that the flow was a circulation about the dorsal-ventral axis from anterior to left \citep{Kreiling07,Supatto08}. However \citeauthor{Kreiling07} and \citeauthor{Supatto08} report different mechanisms which may be responsible for creating the flow. We have developed a theoretical model of flow in the complex geometry of KV using the regularized Stokeslet boundary integral equation. We evaluated the flow field produced by models based on the cilia position and tilt direction features reported by \citeauthor{Kreiling07} and \citeauthor{Supatto08}, both separately, and in combination.

Observed profiles are most closely fitted by a model that combines the experimental observations of posterior and dorsal tilt, and moreover takes into account increased cilia density at the anterior of the dorsal roof. Analysis of flow in a transverse section suggests that the mixed posterior and dorsal tilt may more closely match experimental observations in medakafish KV than dorsal-only tilt, in particular showing a leftward flow near the dorsal roof and rightward counterflow near the ventral floor. However, medakafish KV is a different biological entity from zebrafish KV, possessing just a single population of cilia on the dorsal roof and no cilia on the ventral floor; future experimental observations may provide more information on the flow field and therefore the requirement for the mixture of dorsal and posterior tilt postulated here. An interesting finding in terms of interpreting experimental observations is that, in the dorsal-anterior corner, the posterior tilt and a dorsal tilt directions are very similar, and may not be easily distinguishable experimentally.

Our computations were restricted to a population of cilia rotating in synchrony with a frequency of $30$ Hz. The latter assumption is not a significant limitation. Reports (see for example \citet{Okabe08}) suggest that rotational frequency lies within a very narrow range, and indeed is close to the estimated value of $30$ Hz. Moreover, the temporal invariance of the Stokes flow equations describing microscale flow entails that fluid velocity is directly proportional to beat frequency; for example that fluid velocity for a beat frequency of $25$ Hz can be calculated directly by multiplying our reported results by $5/6$. The assumption of cilia synchronisation is not likely to affect time-averaged flow properties due to the relative sparsity of the cilia; particle tracking studies however may reveal significant changes to trajectories resulting from non-synchronisation. For example, even changes to cilia phase difference (see for example \citet{Smith07} figure 12) can significantly alter the trajectories of particles released near to cilia.

There are a number of questions that remain unanswered regarding symmetry breaking flow in KV. Organising structures are growing, transient entities and the changes in size, ciliation and flow field during its existence remain to be elucidated; more extensive experimental observations and modelling will provide information on this issue.
The process of how an asymmetric flow is translated into asymmetric development is still unclear. A leftward flow may allow morphogens to be transported to the left allowing a concentration gradient to be established across the left-right axis, subject to the morphogens being bound or rendered inactive before they are returned by the counterflow \citep{Cartwright04}. If such morphogens are enclosed in micron-sized or larger lipoprotein vesicles, as reported in the mouse \citep{Tanaka05}, the effect of diffusion may be relatively small compared with advection; we may predict the movement of such vesicles through Lagrangian particle tracking, or simulations which include finite-sized rigid or deformable vesicles suspended in the fluid. Furthermore, it will be possible to calculate the trajectories of individual particles from a range of initial positions and test for evidence of chaos in the domain, as discussed previously \citep{Smith07,Supatto08} by calculating Lyapunov exponents for a group of particles initially close together \citep{Otto01}. A related question is how finite-sized particles, including vesicles and other particles imaged experimentally, affect the flow field. The competing hypothesis of mechanical sensing by passive cilia or other mechanotranducers may also be investigated using similar approaches.

It is not clear how KV cilia `know' which way to tilt. In the mouse node, the migration of the cilium basal body towards the posterior, combined with the convex cell surface, establishes the posterior tilt \citep{Hashimoto10}. The mechanism responsible for the more complex cilia tilt distribution on the inner surface of KV is unclear. Modelling may be beneficial in helping to elucidate any possible cooperative hydrodynamic effect through which organised cilia orientation may emerge.

This study emphasises the tractability of modelling biological flow problems involving relatively complex geometries without the need for simplifying assumptions such as cilium slenderness and precisely planar/spherical boundaries. Examples of other systems which can be modelled in this more refined way include the mouse ventral node; surface features that have not previously been considered, such as the uneven ciliated surface may be taken into account. Finally, and in keeping with the theme of this special issue, we remark that an `inverted Kupffer's vesicle' bears striking similarities to more frequently-studied objects of biological fluid mechanics modelling, for example \textit{Paramecia}, \textit{Chlamydomonas} \citep{Ishikawa07,Pedley92} and \textit{Volvox} \citep{Drescher09}, and therefore the theory in this paper may be generalised to the study of their swimming behaviour.

\section*{Acknowledgements}
AAS and TDJ acknowledge studentships from EPSRC. DJS acknowledges funding from Birmingham Science City. The computations described in this paper were performed using the University of Birmingham's  BlueBEAR HPC service, which was purchased through HEFCE SRIF-3 funds. See \underline{http://www.bear.bham.ac.uk} for more details.

JRB acknowledges the significant contributions Professor Tim Pedley has made to the development of his career over the last 40 years.
The authors gratefully acknowledge comments from the anonymous referees.

\section*{Appendix: numerical implementation}

Integrals were evaluated using standard numerical integration rules. Numerical tests showed that a $3\times3$ Gauss-Legendre quadrature \citep{Abramowitz64} for the curved quadrilateral elements and a $3$ point Fekete rule \citep{Taylor01} for the curved triangular elements provided satisfactory accuracy, provided the distance between the evaluation point and the element centroid is greater than $\alpha\times$(element length), where $\alpha = \sqrt{21}$ for the quadrilateral elements, $\alpha = \sqrt{40}$ for the triangular elements and element length was taken to be the longest side of a triangular element and the greatest distance between opposing corner points of quadrilateral elements. The remaining integrals were evaluated using an adaptive integration routine; briefly, integration results were compared for different rules until the difference between successive rules converged to within a specified tolerance of $10^{-7}$. These values of $\alpha$ were decided upon by calculating each integral for one timestep using our adaptive integration routine everywhere and establishing at which distance was a suitable cut off to switch between a low-order integration rule and using the adaptive integration routine. A tolerance of $10^{-5}$ was sufficient for computations; results were reported for computations performed with a tolerance of $10^{-7}$. A regularization of $\epsilon=0.01\mu$m was used.

Meshes contained $N = 6252$ elements, resulting in $3N$ scalar degrees of freedom for the stress $f_j[m]$ at each timestep. Results were reported with $N_t=60$ timesteps. Because there is no explicit time-dependence in the Stokes flow equations, calculations at each timestep were independent. Restarted GMRES was used to solve the linear system (FORTRAN library routines F11BDF/BEF/BFF, F11XAF, F11DBF, Numerical Algorithms Group, Oxford).

%To solve the discrete approximations, $q_i[n]$, we first write equation \eqref{eq:disc-sys} as a matrix equation, $A_{ij}X_i = B_j$ where $A_{ij} = \int_{S[n]}{K_{ij}(\bm{x},\bm{y})\mathrm{d}S_{\bm{x}}}$, $X_i = q_i[n]$ and $B_j = u_j(\bm{y},t)$, we then impose a velocity on the right hand side as described in section \ref{subsec:mesh} and complete the matrix $A_{ij}$.

\bibliographystyle{jfm2}
\bibliography{myrefs}

\end{document}